\documentclass[twocolumn]{aastex63}
\usepackage{mathtools}
\usepackage{graphicx}
\usepackage{physics}
\usepackage{xcolor}
\usepackage{hyperref}

\submitjournal{ApJ}

\shorttitle{The 21\,cm--kSZ--kSZ Bispectrum during the EoR}
\shortauthors{La Plante et al.}

\begin{document}

\title{The 21\,cm--kSZ--kSZ Bispectrum during the Epoch of Reionization}


\author[0000-0002-4693-0102]{Paul La Plante}
\affiliation{Astronomy Department, University of California,
Berkeley, CA 94720 USA; \href{mailto:plaplant@berkeley.edu}{plaplant@berkeley.edu}}
\affiliation{Berkeley Center for Cosmological Physics, University of California,
Berkeley, CA 94720 USA}
\affiliation{Center for Particle Cosmology,
Department of Physics and Astronomy, University of Pennsylvania,
Philadelphia, PA 19104 USA}
\author[0000-0002-3950-9598]{Adam Lidz}
\affiliation{Center for Particle Cosmology,
Department of Physics and Astronomy, University of Pennsylvania,
Philadelphia, PA 19104 USA}
\author[0000-0002-4810-666X]{James Aguirre}
\affiliation{Center for Particle Cosmology,
Department of Physics and Astronomy, University of Pennsylvania,
Philadelphia, PA 19104 USA}
\author[0000-0001-6744-5328]{Saul Kohn}
\affiliation{Center for Particle Cosmology,
Department of Physics and Astronomy, University of Pennsylvania,
Philadelphia, PA 19104 USA}

\begin{abstract}
  Current- and next-generation radio interferometers such as the Hydrogen Epoch
  of Reionization Array (HERA) and Square Kilometre Array (SKA) are projected to
  measure the 21\,cm auto-power spectrum from the Epoch of Reionization
  (EoR). Another observational signal of this era is the kinetic
  Sunyaev--Zel'dovich (kSZ) signal in the cosmic microwave background (CMB),
  which will be observed by the upcoming Simons Observatory (SO) and CMB-S4
  experiments. The 21\,cm signal and the contribution to the kSZ from the EoR
  are expected to be anticorrelated. However, the na\"\i ve cross-correlation
  between the kSZ and 21\, cm maps suffers from significant cancellation. We
  present here an investigation of the 21\,cm--kSZ--kSZ bispectrum, which should
  not suffer the same cancellation as the simple two-point cross-correlation. We
  show that there is a significant and nonvanishing signal that is sensitive to
  the reionization history. In the absence of foreground contamination, we
  forecast that this signal is detectable at high statistical significance with
  HERA and SO. However, the bispectrum we study suffers from the fact that the
  kSZ signal is sensitive only to Fourier modes with long-wavelength
  line-of-sight components, which are generally lost in the 21\,cm data sets
  owing to foreground contamination. We discuss possible strategies for
  alleviating this contamination, including an alternative four-point statistic
  that may help circumvent this issue.
\end{abstract}


\section{Introduction}
During the Epoch of Reionization (EoR), the intergalactic medium (IGM) underwent
a large-scale phase change, transitioning from neutral to ionized gas. Neutral
hydrogen gas in the IGM may be observed in emission or absorption against the
background cosmic microwave background (CMB) at a rest wavelength of
$\lambda = 21$\,cm. This signal is being pursued observationally through radio
interferometer telescopes such as the Hydrogen Epoch of Reionization Array
(HERA\footnote{\url{https://reionization.org}}), the Low Frequency Array
(LOFAR\footnote{\url{https://www.lofar.org}}), and the Square Kilometre Array
(SKA\footnote{\url{https://www.skatelescope.org}}). These arrays seek to measure
statistical correlations in the 21\,cm signal from the EoR at a high
signal-to-noise ratio (S/N), which would provide insight into the topology of
reionization and yield vital clues as to the astrophysical sources responsible
for reionization.

In addition to the 21\,cm signal from the EoR, theoretical calculations suggest
there should exist a complementary signature in the kinetic Sunyaev--Zel'dovich
(kSZ) effect. The kSZ is a secondary distortion of the CMB, in which CMB photons
inverse-Compton scatter off of free electrons and receive Doppler shifts owing
to the bulk motion of these electrons. The kSZ signal was first detected by the
Atacama Cosmology Telescope (ACT;\footnote{\url{https://act.princeton.edu}}
\citealt{hand_etal2012}), and since then has been measured by the South Pole
Telescope (SPT;\footnote{\url{https://pole.uchicago.edu/}}
\citealt{soergel_etal2016}) and the Planck mission
\citep{planck_ksz}. The kSZ has a component due to the contribution of the
ionized gas in the post-reionization universe, which is dominated by galaxy
clusters with large peculiar velocities relative to the CMB. There is also the
so-called ``patchy'' contribution, considered here, that is due to the fact that
reionization is an inhomogeneous process that took a significant amount of
cosmic time to complete. Specifically, spatial variations in the timing of
reionization imprint CMB anisotropies on angular scales spanning tens of
arcminutes. These provide important information regarding when and how
reionization occurs. Recently, the SPT-SZ and SPTpol missions reported a
detection of the kSZ signal at 3$\sigma$ significance, with a value of
$D_{3000}^\mathrm{kSZ} = 3.0 \pm 1.0$ $\mu$K$^2$
\citep{reichardt_etal2020}. Measuring the kSZ signal is one of the science goals
of the upcoming Simons Observatory
(SO\footnote{\url{https://simonsobservatory.org}}) and
CMB-S4\footnote{\url{https://cmb-s4.org/}} experiments. During the EoR, sources
of UV photons emit radiation into the IGM and ionize neutral hydrogen as it
propagates. The ionized gas no longer emits 21\,cm radiation, but instead can
generate a nonzero kSZ signal.

As with any astrophysical signal, systematic errors can be mitigated by
performing measurements of cross-correlations, in which the signal from two
datasets can be analyzed to understand what statistical trends can be seen in
both datasets. In the case of the 21\,cm and kSZ signals from the EoR, one would
naively expect a signal to be present, because the same astrophysical sources---
namely, objects producing UV photons---give rise to both signals. However, the
kSZ signal can be either positive or negative depending on whether the free
electrons are moving toward or away from us. Due to the large-scale isotropy and
homogeneity of the universe, the simple cross-correlation measurement between
the kSZ signal and the 21\,cm signal is expected to suffer from significant
cancellation on scales where the kSZ signal is larger than the primary CMB
signal ($\ell \gtrsim 3000$)\footnote{As discussed further in
  Sec.~\ref{sec:bspec_background}, this cancellation is partly avoided on large
  scales, but the signal is hard to measure in practice
  \citep{alvarez_etal2006,adshead_furlanetto2008,alvarez2016,ma_etal2018}.}. Such
a cancellation is common with other tracers that cross-correlate with the kSZ
signal, such as weak lensing or galaxy surveys. One approach in such situations
is simply to square the kSZ signal, taking care to filter out the primary CMB
component and other sources of unwanted noise
\citep{dore_etal2004,hill_etal2016}. The kSZ$^2$-21\,cm two-point
cross-correlation was explored in \citet{ma_etal2018}. Here, we pursue the more
general approach of computing a cross-bispectrum between two kSZ fields and the
21\,cm field. The statistic in \citet{ma_etal2018} is closely related to ours,
and may be expressed as an integral over the quantity we consider. Although our
statistic is more complex and computationally demanding, it potentially contains
more information: we explicitly consider the dependence over a broad range of
triangles in harmonic space. This may also provide greater flexibility and
control over systematics.

To model the 21\,cm--kSZ--kSZ bispectrum, we employ semi-numeric simulations of
reionization. The two fields are generated from the same realization of
reionization, which allows for a self-consistent study of the two fields. The
semi-numeric simulations are fast yet incorporate important nonlinearities in
the density, ioniziation fraction, 21\,cm, and kSZ fields. The strong
fluctuations in these fields during reionization generally necessitate modeling
beyond linear perturbation theory.  The rest of the paper is outlined as
follows. In Sec.~\ref{sec:bspec}, we review the theory of the kSZ and 21\,cm
signals. In Sec.~\ref{sec:methods}, we discuss the methods by which we generate
the 21\,cm field and kSZ field, and compute the bispectrum. In
Sec.~\ref{sec:results}, we show the results of our theoretical calculation, and
present a qualitative picture for explaining the results. In Sec.~\ref{sec:snr},
we discuss detectability as it relates to the sample variance as well as
instrumental noise estimates for upcoming experiments. In
Sec.~\ref{sec:conclusion}, we conclude and discuss future directions. Throughout
the text, we assume a $\Lambda$CDM cosmology with parameters consistent with the
Planck 2018 results \citep{planck2018}.

\section{The 21\,cm--kSZ--kSZ Bispectrum}
\label{sec:bspec}

\subsection{The 21\,cm Field}
\label{sec:21cm}

The 21\,cm signal is generated by neutral hydrogen in the IGM. The signal itself
is a brightness temperature that depends on the spin temperature of the hydrogen
gas. The 21\,cm signal can be expressed as \citep{madau_etal1997}:
\begin{equation}
\delta T_b(\vb{r},z) = T_0(z) [1 + \delta_m(\vb{r})] [1 - x_i(\vb{r})],
\label{eqn:t21}
\end{equation}
where $\delta_m$ is the matter overdensity, and $x_i$ is the ionization fraction
of the gas ($x_i = 1$ is totally ionized gas, $x_i = 0$ is totally
neutral). $T_0(z)$ is
\begin{multline}
T_0(z) = 26\qty(\frac{T_S - T_\gamma}{T_S})\qty(\frac{\Omega_b h^2}{0.022}) \\
  \times \qty[\qty(\frac{0.143}{\Omega_m h^2})\qty(\frac{1+z}{10})]^{\frac{1}{2}} \, \mathrm{mK}, \label{eqn:t0}
\end{multline}
where $T_S$ is the spin temperature of neutral hydrogen, and $T_\gamma$ is the
temperature of the CMB. We assume that $T_S$ is coupled to $T_\mathrm{gas}$
throughout the entire IGM and that the gas temperature is globally much larger
than the CMB temperature. This assumption is valid once both a sufficient
Ly$\alpha$ background has built up to couple the spin temperature to the gas
temperature and once early X-rays or other sources provide the needed heat
input. This may occur once the globally averaged ionization fraction
$\ev{x_i} \gtrsim 0.25$ \citep{santos_etal2008}, although significant
uncertainties remain regarding the onset of the Ly$\alpha$ background and the
timing of early X-ray heating
\citep{pritchard_furlanetto2007,mirocha2014,eide_etal2018}. As shown in
\citet{greig_mesinger2018}, incorrectly assuming spin temperature saturation can
bias the recovery of semi-analytic model parameters. In the application at hand,
we are interested in understanding the relationship between the 21\,cm and kSZ
signals during the central portion of the EoR and at late times, where the
assumption of spin temperature saturation is likely justified. As such, the
results at high redshift during the pre-reionization epoch ($z \gtrsim 10$ for
the fiducial model presented here) may be inaccurate, though the primary results
pertain to epochs when the assumption is well justified.

The 21\,cm signal is a spectral line, and so in principle, the signals from
different redshifts can be detected independently, giving full 3D tomographic
information of the EoR. Accordingly, the 21\,cm field can provide valuable
information about the entire process of reionization. A major scientific goal of
both HERA and the SKA is to generate maps of the EoR for redshifts
$z \lesssim 12$, with sufficient fidelity to observe the formation of ionized
regions surrounding galaxies. These maps will themselves provide rich insight
into the astrophysical details of the first luminous sources, and are also prime
candidates for performing cross-correlation analysis. In addition to the kSZ
signal discussed below, other intensity mapping tracers such as [C~\textsc{ii}]
\citep{beane_lidz2018} can provide significant insight into the EoR. The
information gleaned from cross-correlation studies can provide important
cross-checks to that from the 21\,cm auto-power spectrum, and may avoid some of
the systematic errors associated with such measurements.

\subsection{The kSZ Field}
\label{sec:ksz}

The kSZ effect is an integrated line-of-sight effect that can be observed in
maps of the CMB. In a direction toward the CMB $\vu{n}$, the kSZ effect can be
expressed as an integral along the line of sight \citep{sunyaev_zeldovich1972}:
\begin{align}
\frac{\Delta T (\vu{n})}{T_\mathrm{CMB}} &= - \frac{\sigma_T}{c} \int \dd{l} n_e e^{-\tau(l)} \vb{v} \cdot \vu{n} \notag \\
&= - \int \dd{\chi} g(\chi) e^{-\tau(\chi)} \vb{q} \cdot \vu{n}
\label{eqn:ksz}
\end{align}
where $\sigma_T$ is the Thomson cross section, $c$ is the speed of light, $l$ is
the proper distance along the line of sight, $n_e$ is the local (proper)
electron density, $\tau = \sigma_T n_e l$ is the local optical depth, and
$\vb{v}$ is the peculiar velocity of the ionized electrons. In the second line,
we have transitioned to co-moving coordinates $\chi$ and introduced the kSZ
visibility function $g(\chi)$, defined as \citep{alvarez2016}:
\begin{equation}
g(\chi) = \pdv{\ev{\tau}}{\chi} = \sigma_T n_{e,0} \ev{x_i} (1 +
z)^2,
\label{eqn:wksz}
\end{equation}
where $n_{e,0} = \qty[1 - (4 - N_\mathrm{He})Y/4]\Omega_b\rho_\mathrm{crit}/m_p$
is the mean electron number density. We set the number of helium ionizations per
hydrogen atom $N_\mathrm{He} = 1$, so that helium is singly ionized along with
hydrogen (which is not doubly ionized until significantly later;
\citealt{laplante_etal2017}) and depends on the helium mass fraction $Y$. This
quantity is multiplied by the local electron momentum
$\vb{q} = \vb{v}(1 + \delta_m)x_i/c$ and integrated along the line of sight
$\chi$ to compute the full spectral distortion of the CMB. The kSZ signal as
observed in the CMB contains contributions from the EoR as well as ionized gas
in the post-reionization era. The contribution to the kSZ effect from patchy
reionization has a typical magnitude of $D_\ell = \ell^2 C_\ell / (2\pi) \sim 3$
$\mu$K$^2$ at a scale of $\ell \sim 3000$
\citep{battaglia_etal2013b,alvarez2016}.

Analogously to how the local 21\,cm fluctuation $\delta T_b$ is defined in
Equation~(\ref{eqn:t21}) for all points $\vb{r}$ in 3D space, we define the local kSZ
fluctuation $\delta_q$:
\begin{equation}
\delta_q(\vb{r}) \equiv T_\mathrm{CMB}  [1 + \delta_m(\vb{r})]x_i(\vb{r}) e^{-\tau},
\label{eqn:qksz}
\end{equation}
where $\tau$ is the local optical depth of the volume\footnote{For the
  simulation resolution considered, $e^{-\tau} \approx 1$, so the inclusion of
  the optical depth does not significantly affect the calculation.}. Note that
there is no explicit dependence on the local velocity field in the definition of
this field, and we include the temperature factor $T_\mathrm{CMB}$ so that it
has temperature units associated with the fluctuation, which is useful for
comparing with power spectra seen in the literature. When constructing
estimators involving two kSZ fields such as the power spectrum, we assume that
the velocity field is coherent on spatial scales much larger than the ones of
interest here ($\ell \gtrsim 3000$; \citealt{mesinger_etal2012}). Accordingly,
we replace the velocity term with $v_\mathrm{RMS}^2 / 3$, where $v_\mathrm{RMS}$
is the rms of the peculiar velocities. For example, the Limber approximation
\citep{limber1953,kaiser1992} for estimating the $C_\ell$ spectrum of the kSZ
field can be expressed as:
\begin{equation}
C_{\ell,\mathrm{kSZ}}(\ell) \approx \frac{1}{3c^2} \int \frac{\dd{\chi}}{\chi^2} v_\mathrm{RMS}^2(\chi) g^2(\chi) P_{qq}(\ell/\chi),
\label{eqn:clksz}
\end{equation}
where $P_{qq}$ is the 3D power spectrum computed from the $\delta_q$ field
defined in Equation~(\ref{eqn:qksz}). This approximation is expected to break
down on large scales corresponding to those where the velocity flows are
coherent (typically $\sim$100 Mpc).  We have verified that the $C_\ell$ spectrum
computed in this fashion and directly from maps of the kSZ signal such as those
in Figure~\ref{fig:maps} shows qualitatively good agreement at $\ell \sim 3000$.

\subsection{The Bispectrum}
\label{sec:bspec_background}

We define the Fourier transform as:
\begin{equation}
\tilde{\delta}(\vb{k}) = \int \frac{\dd[3]{r}}{(2\pi)^3} \delta(\vb{r}) e^{-2\pi i \vb{k} \cdot \vb{r}}.
\label{eqn:ft}
\end{equation}
Using this convention, the bispectrum $B(k_1,k_2,k_3)$ is
\citep{scoccimarro_etal1998}:
\begin{multline}
\ev{\tilde{\delta}(\vb{k}_1)\tilde{\delta}(\vb{k}_2)\tilde{\delta}(\vb{k}_3)} = \\
(2\pi)^3 \delta_D(\vb{k}_1 + \vb{k_2} + \vb{k_3}) B(k_1, k_2, k_3),
\label{eqn:bspec}
\end{multline}
where $\delta_D$ is the Dirac delta function, which guarantees that the Fourier
modes chosen form a closed triangle. Analogously to the power spectrum, one can
compute the auto-bispectrum, where all three constituent fields in
Equation~(\ref{eqn:bspec}) are the same quantity. The auto-bispectrum of the
21\,cm field during the EoR has recently been studied
\citep{shimabukuro_etal2016,majumdar_etal2018,giri_etal2019}, which reveals
interesting non-Gaussian correlations present in the 21\,cm field. In this work,
we examine the cross-bispectrum between two kSZ fields and one 21\,cm
field. Specifically, the quantity of interest is $B_{\mathrm{21cm},q,q}$,
defined as:
\begin{multline}
\ev{\tilde{\delta T}_b(\vb{k}_1)\tilde{\delta}_q(\vb{k}_2)\tilde{\delta}_q(\vb{k}_3)} = \\
(2\pi)^3 \delta_D(\vb{k}_1 + \vb{k_2} + \vb{k_3}) B_{\mathrm{21cm},q,q}(k_1, k_2, k_3).
\label{eqn:tkk_bspec}
\end{multline}
Note that the 3D bispectrum defined in this fashion has units of
$\qty[\mathrm{mK}\mu\mathrm{K}^2(h^{-1}\mathrm{Mpc})^6]$. Unlike the power
spectrum, the bispectrum can take on positive or negative values: positive
values represent correlation between the constituent fields for the modes
probed, and negative values represent anticorrelation.

Previous work has explored the two-point cross-power spectrum between the kSZ
field and 21\,cm field during the EoR. \citet{alvarez_etal2006} demonstrated a
substantial cross-correlation on degree scales ($\ell \sim 100$) between the
fields due to the so-called Doppler term. The physical explanation for this
cross-correlation comes from considering a matter overdensity, and recognizing
that there is less ionized gas on the far side of this overdensity from the
observer (corresponding to higher redshift) and falling toward the observer, and
more ionized gas on the near side of the overdensity and falling away from the
observer. Because these regions correspond to a decrease in the expected kSZ
signal, as well as the 21\,cm signal, the overall cross-correlation is large and
positive (though far below the level of the primary CMB power spectrum).
However, this cross-correlation falls off sharply as a function of $\ell$, and
is essentially zero for scales corresponding to ionized bubbles from the EoR
($\chi \sim 10$ $h^{-1}$Mpc, $\ell \sim 3000$) because individual bubbles may be
moving toward or away from the observer with equal probability, leading to large
cancellation. \citet{alvarez2016} built upon and extended this previous result
to explore auto- and cross-correlations of all relevant combinations of the
components contributing to the overall signal. \citet{ma_etal2018} also
investigated the cross-correlation between the kSZ field and 21\,cm fields, and
explored the signal coming from squaring the kSZ field in real space in an
attempt to avoid the cancellation of the velocity. This work explicitly
considers the 21\,cm--kSZ--kSZ bispectrum. By using two instances of the kSZ field
in the bispectrum calculation, the cancellation of the line-of-sight velocity in
the kSZ field can be mitigated.

Due to the fact that the kSZ signal is fundamentally 2D, the bispectrum analog
of the Limber approximation can be used to compute the 21\,cm--kSZ--kSZ bispectrum
$\mathcal{B}_\mathrm{21cm,kSZ,kSZ}$ \citep{buchalter_etal2000}:
\begin{multline}
\mathcal{B}_{\mathrm{21cm,kSZ,kSZ}}(\ell_1, \ell_2, \ell_3) = \frac{1}{3c^2} \int \frac{\dd{\chi}}{\chi^4} W_\mathrm{21 cm} (\chi) g^2 (\chi) \\
\times  v_\mathrm{RMS}^2(\chi) B_{\mathrm{21 cm},q,q}(\ell_1/\chi, \ell_2/\chi, \ell_3/\chi),
\label{eqn:limber}
\end{multline}
where $\chi$ is the co-moving distance along the line of sight,
$W_\mathrm{21cm}(\chi)$ is the window function associated with the 21\,cm field,
and $g(\chi)$ is the kSZ visibility function defined in
Equation~(\ref{eqn:wksz}). This equation is analogous to the original bispectrum
definition in Equation~(\ref{eqn:tkk_bspec}), but defined in $\ell$-space rather
than $k$-space. As discussed in Sec.~\ref{sec:ksz}, we use the 3D analog of the
kSZ field defined in Equation~\ref{eqn:qksz} when computing
$B_{\mathrm{21cm},q,q}$. We also make the substitution of using
$v_\mathrm{RMS}^2$ in lieu of the peculiar velocity, as in
Equation~(\ref{eqn:clksz}). The 21\,cm window function quantifies the
contribution of each segment along the co-moving line of sight $\dd{\chi}$ to the
overall result, and is normalized such that
$\int\dd{\chi} W_\mathrm{21cm}(\chi) = 1$. There is some more flexibility in
choosing the window function, though in practice it is determined by the
observational strategy. In the results below, we show the impact that the choice
of the 21\,cm window function has on the results. For further discussion, see
Sec.~\ref{sec:2dbspec}.

\begin{figure}
  \centering
  \includegraphics[width=0.45\textwidth]{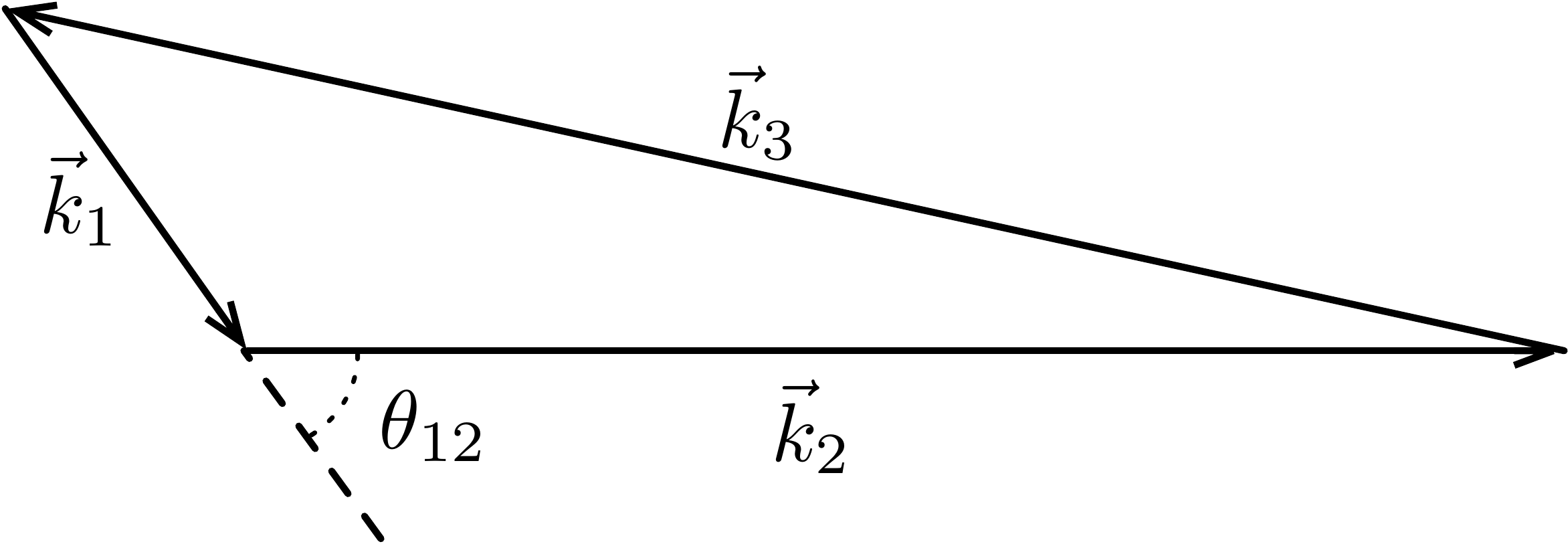}
  \caption{An example of a squeezed bispectrum triangle. The Dirac delta
    function in Equation~(\ref{eqn:tkk_bspec}) ensures that the triangle is
    closed. The length $k_1 \ll k_2,k_3$, meaning that $\vb{k}_1$ is a probe of
    large-scale structure while $\vb{k}_2$ and $\vb{k}_3$ probe smaller
    scales. The triangle is defined by the lengths $k_1$ and $k_2$ and the angle
    between them $\theta_{12}$. As mentioned in Sec.~\ref{sec:bspec}, we
    primarily discuss results from the squeezed-triangle limit.}
  \label{fig:triangle}
\end{figure}

In the results that follow, we examine the squeezed-triangle limit of the
bispectrum. The Dirac delta function in Equation~(\ref{eqn:tkk_bspec}) requires
that the three Fourier modes chosen form a closed triangle. We further restrict
ourselves to the length of one leg of the triangle being significantly shorter
than the other two. In particular, we choose the short leg of the triangle
$\vb{k}_1$ to correspond to the 21\,cm field, and the other two to correspond to
the kSZ field. This choice is motivated primarily by the relevant scales of
upcoming experiments: HERA is projected to observe the 21\,cm field at scales of
$90 \lesssim \ell \lesssim 1000$ (transverse modes $0.014$ $h^{-1}$Mpc
$\lesssim k \lesssim 0.15$ $h^{-1}$Mpc at $z=8$), whereas SO is projected to
observe the kSZ signal in the CMB at scales of
$3000 \lesssim \ell \lesssim 6000$ ($0.49$ $h^{-1}$Mpc
$\lesssim k \lesssim 0.98$ $h^{-1}$Mpc at $z=8$). This difference in scales
naturally leads to a squeezed-triangle configuration. Furthermore, the
squeezed-triangle limit can be thought of as a conditional measurement: given a
large-scale region of excess 21\,cm brightness temperature, it quantifies
whether the small-scale kSZ power spectrum is larger or smaller than in regions
of average 21\,cm brightness temperature.

Figure~\ref{fig:triangle} shows a sample squeezed triangle. $\vb{k}_1$
corresponds to the 21\,cm field, and $\vb{k}_2$ and $\vb{k}_3$ correspond to the
kSZ field. The angle $\theta_{12}$ between the vectors $\vb{k}_1$ and $\vb{k}_2$
is defined as: \begin{equation} \theta_{12} \equiv \cos^{-1}
\left[\left(\vb{k}_1 \cdot \vb{k}_2\right) / (k_1 k_2)\right] \label{eqn:theta}
\end{equation} Statistically, the bispectrum is completely characterized by the
lengths of two legs $k_1$, $k_2$ and the angle between them $\theta_{12}$. Using
Equation~(\ref{eqn:theta}), it is therefore possible to interchange between
specifying the bispectrum using $\{k_1,k_2,k_3\}$ and $\{k_1, k_2,
\theta_{12}\}$.

\section{Numerical Methods}
\label{sec:methods}

\subsection{Reionization Model}
\label{sec:zreion}

To explore the expected level of the 21\,cm--kSZ--kSZ bispectrum from the EoR, we
use the output of $N$-body simulations combined with a semi-numeric model of
reionization. We use a P$^3$M algorithm described in \citet{trac_etal2015} to
generate matter overdensity and velocity fields. The simulation tracks 2048$^3$
dark matter particles in a volume of 2 $h^{-1}$Gpc on a side. To generate the
ionization field $x_i(\vb{r}, z)$ for every point in the volume, we use the
semi-numeric model introduced in \citet{battaglia_etal2013a}. This model has
already been applied to 21\,cm studies in \citet{laplante_etal2014} and
\citet{laplante_ntampaka2019}, and to kSZ studies in \citet{battaglia_etal2013b}
and \citet{natarajan_etal2013}. The starting point for this model is to consider
the redshift at which different regions of the universe become highly ionized
(with ionization fraction $x_i \sim 1$). This is used to define a local redshift
of reionization field $z_\mathrm{re}(\vb{r})$, and the fractional fluctuations
in this quantity $\delta_z(\vb{r})$:
\begin{equation}
\delta_z(\vb{r}) \equiv \frac{\qty[z_\mathrm{re}(\vb{r}) + 1] - \qty[\bar{z} + 1]}{\bar{z} + 1},
\label{eqn:deltaz}
\end{equation}
where $\bar{z}$ is the mean value or reionization. The reionization field
$\delta_z$ is assumed to be a biased tracer of the dark matter overdensity field
on large scales ($\geq 1$ $h^{-1}$Mpc). To quantify the precise relationship
between the fields, a bias parameter $b_{zm}(k)$ is introduced:
\begin{equation}
b_{zm}^2(k) \equiv \frac{\ev{\delta^*_z \delta_z}_k}{\ev{\delta^*_m \delta_m}_k} = \frac{P_{zz}(k)}{P_{mm}(k)}.
\end{equation}
We parameterize the bias parameter $b_{zm}(k)$ as a function of spherical
wavenumber $k$ in the following way:
\begin{equation}
b_{zm}(k) = \frac{b_0}{\qty(1 + \frac{k}{k_0})^\alpha}.
\label{eqn:bias}
\end{equation}
We use the value of $b_0 = 1/\delta_c = 0.593$.  The reionization field for a
given density field is then completely specified by the three values of the
parameters $\{\bar{z}, \alpha, k_0\}$. The parameter $\bar{z}$ is defined in
Eqn.~(\ref{eqn:deltaz}), which determines the midpoint of reionization. The
parameters $k_0$ and $\alpha$ are defined in Eqn.~(\ref{eqn:bias}) and determine
the duration. For this study, we use the values of $\bar{z} = 8$,
$\alpha = 0.564$, and $k_0 = 0.185$ Mpc$^{-1} h$ to represent the fiducial
reionization history. Once the redshift of reionization field
$z_\mathrm{re}(\vb{r})$ has been generated, the ionization field $x_i(\vb{r},z)$
can be computed for a given redshift $z$: if $z_\mathrm{re}(\vb{r})$ is greater
than the target redshift, the cell is treated as being ionized ($x_i=1$), and
neutral ($x_i=0$) otherwise.

To understand how the predicted quantities change as a function of the
ionization history, we adjust the values of $\alpha$ and $k_0$ such that shorter
and longer histories are produced relative to our fiducial history, while still
centered at $\bar{z} = 8$. We also produce a realization that uses the same
$\alpha$ and $k_0$ parameters as the fiducial reionization scenario, but with
$\bar{z} = 10$. These realizations allow us to explore the relationship between
the reionization history and the observed cross-correlation between the two
signals.

\subsection{kSZ and 21\,cm Maps}
\label{sec:maps}

\begin{figure*}
  \centering
  \mbox{}
  \includegraphics[width=0.49\textwidth]{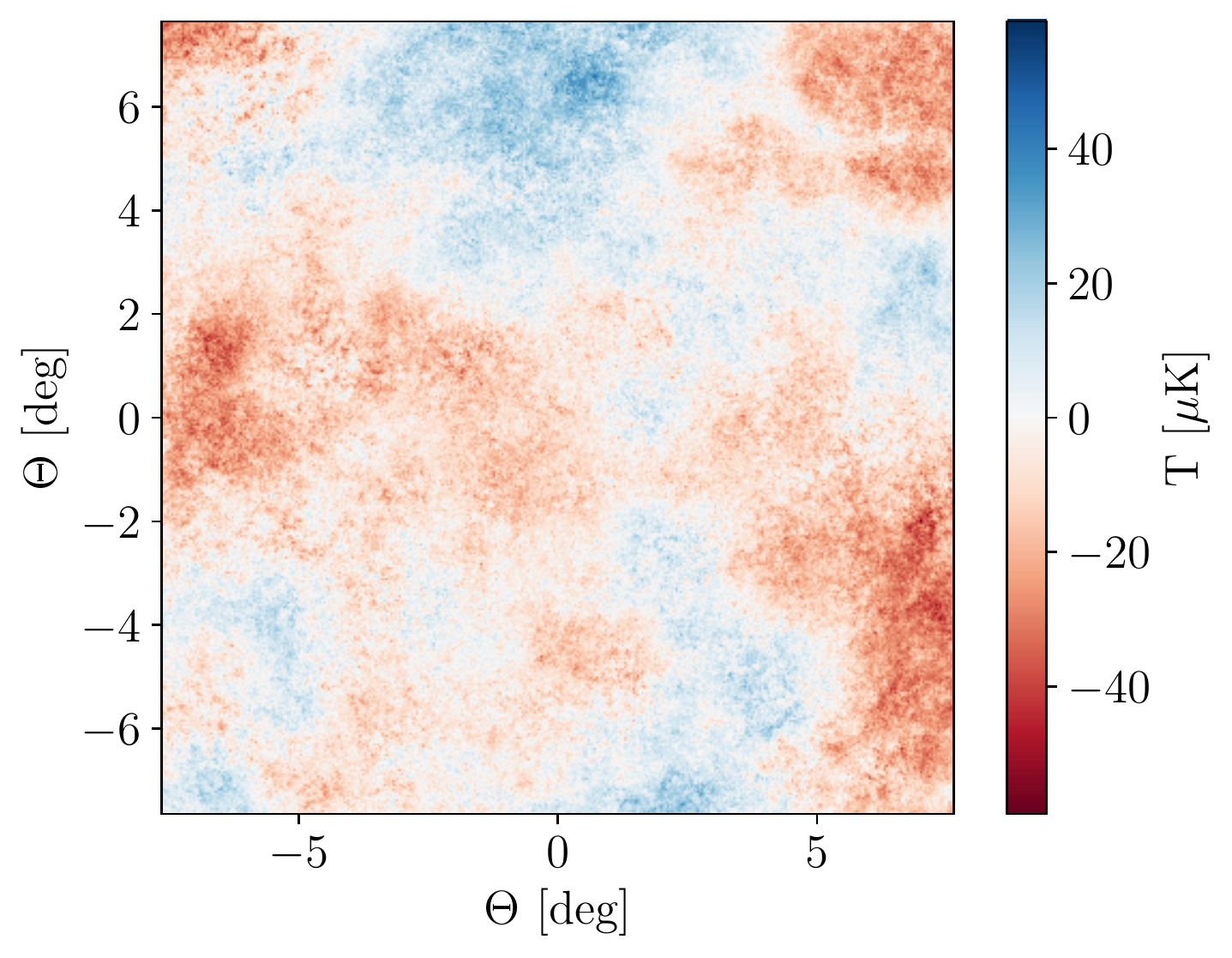}\hfill
  \includegraphics[width=0.49\textwidth]{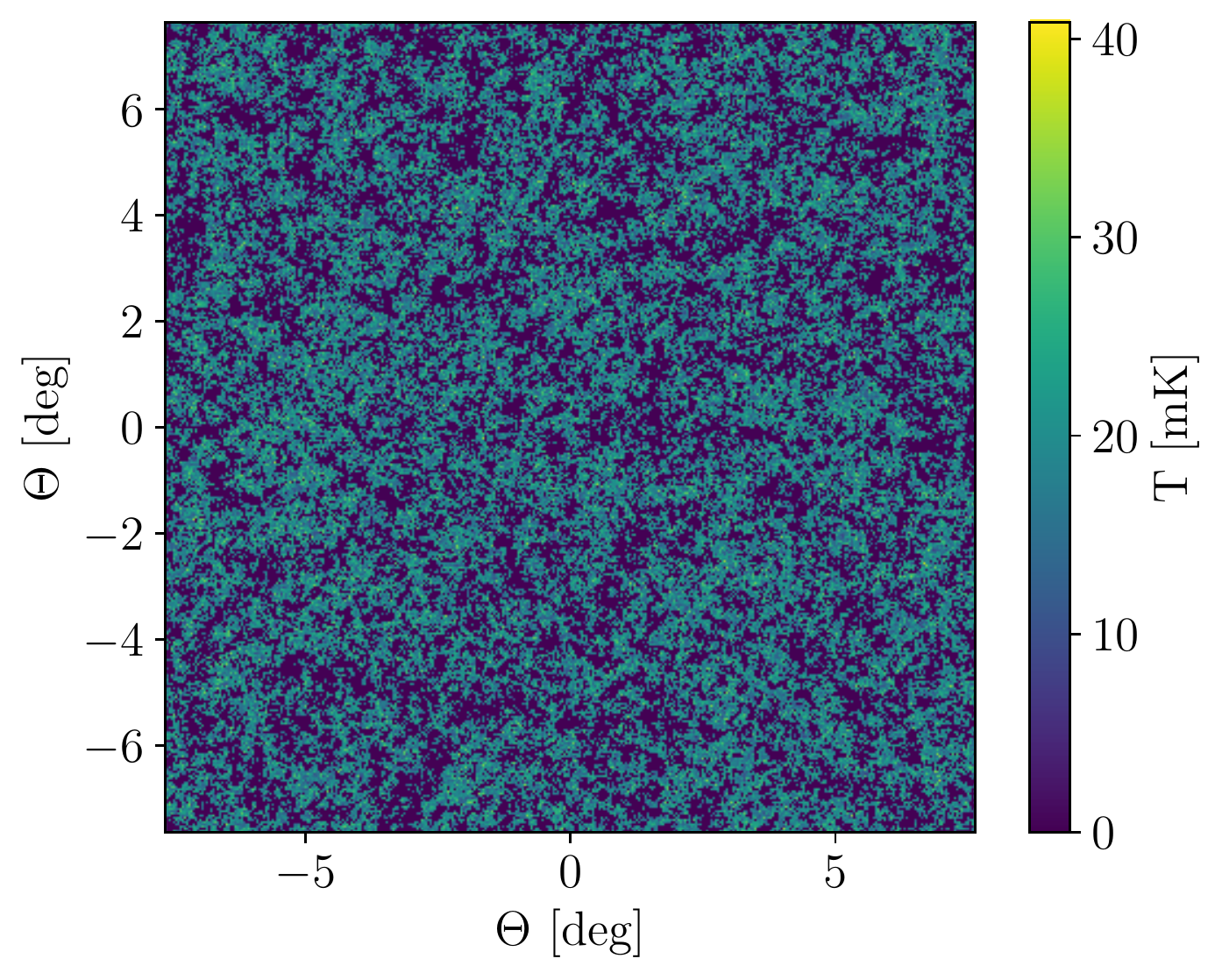}\\
  \includegraphics[width=0.49\textwidth]{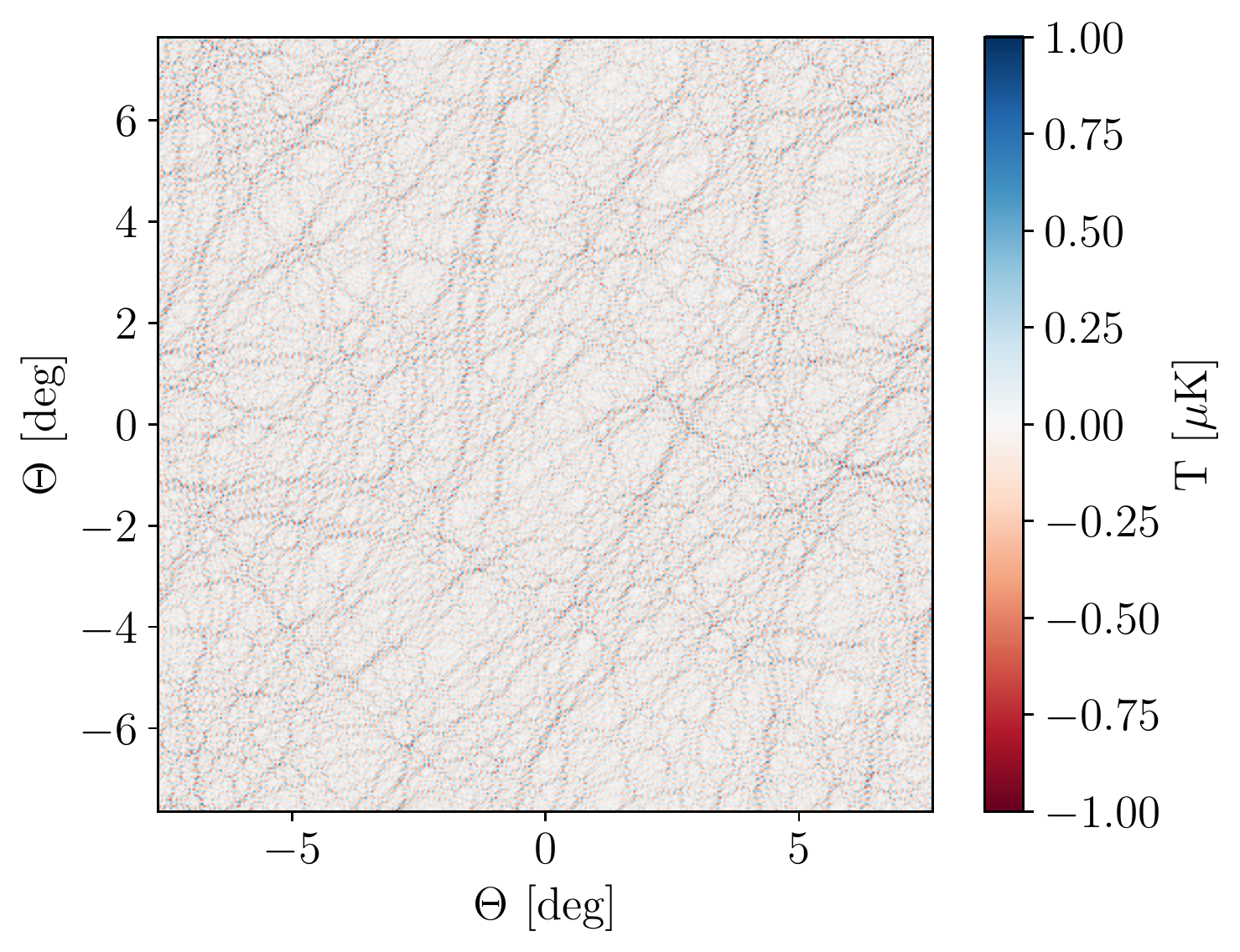}\hfill
  \includegraphics[width=0.49\textwidth]{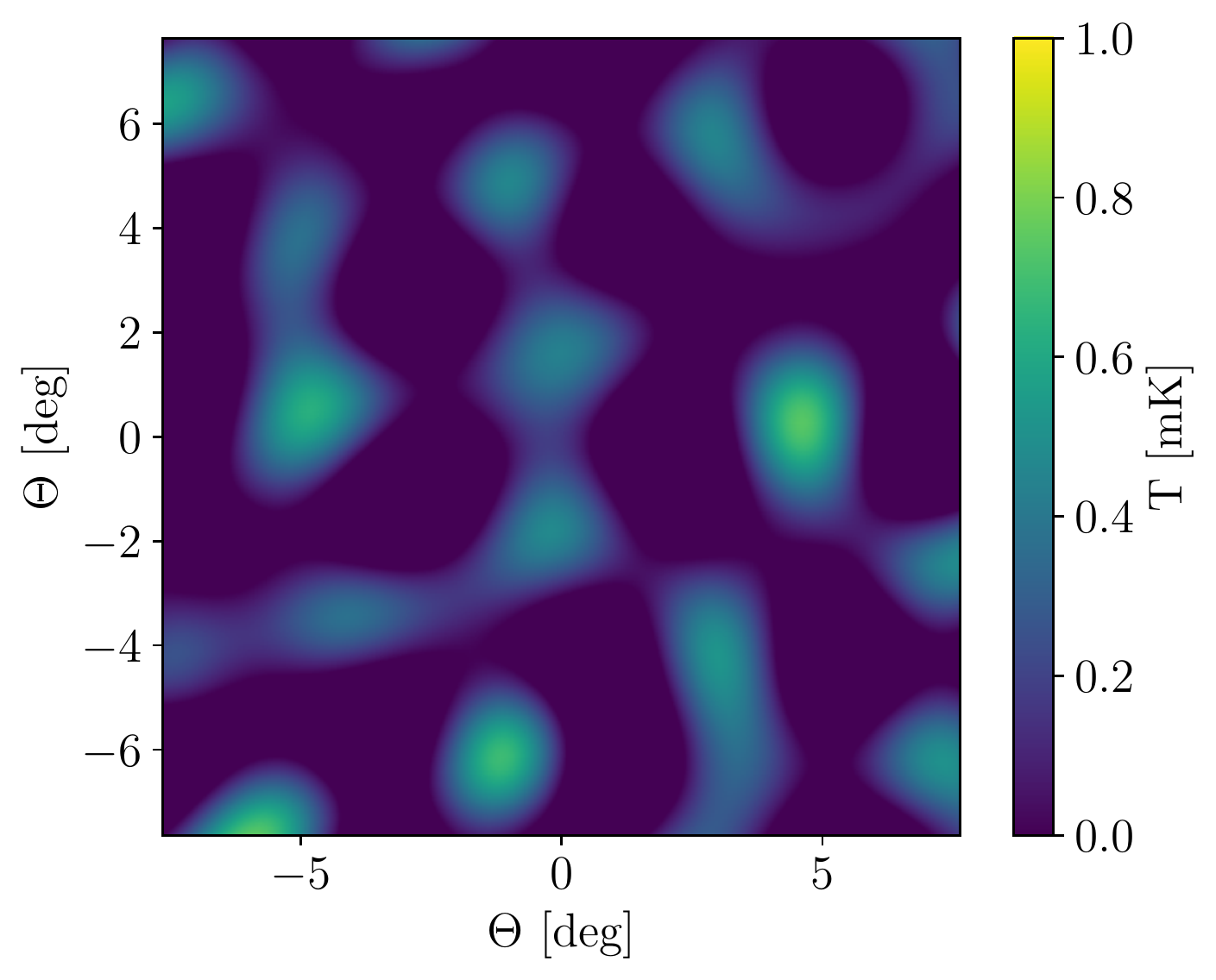}
  \mbox{}
  \caption{Top left: the kSZ map generated from the simulations described in
    Sec.~\ref{sec:methods}. Top right: the 21\,cm map generated from the same
    simulation at $z=8$, near the midpoint of reionization. Bottom left: the kSZ
    map filtered with a Gaussian window with a mean of $\ell = 3000$ and
    standard deviation $\sigma_\ell = 100$. Bottom right: the 21\,cm map
    filtered with a Gaussian window with a mean of $\ell = 90$ and standard
    deviation $\sigma_\ell = 30$. Features at these scales are the ones of
    interest in the cross-correlation statistic we describe below: we examine
    the extent to which large-scale features in the 21\,cm map are correlated
    with small-scale features in the kSZ map. Although these slices to not
    reveal obvious correlations through inspection, there is in fact a
    measurable signal, as we will quantify below in Sec.~\ref{sec:results}.}
  \label{fig:maps}
\end{figure*}

With the ionization field generated in this manner, in principle it is possible
to compute the 3D 21\,cm field using Equation~(\ref{eqn:t21}) and the kSZ map
using Equation~(\ref{eqn:ksz}). One would then define a 21\,cm window function
$W_{21\mathrm{cm}}(\chi)$ to produce a projected, 2D 21\,cm field. At this
point, one could compute the bispectrum using traditional three-point estimators
\citep{spergel_goldberg1999,coulton_etal2018}.  However, the squeezed triangle
suffers from relatively low S/N due to the comparatively few number of small-$k$
(large-scale) modes available even in a large cosmological volume. To overcome
the low S/N, many independent self-consistent realizations of the kSZ map and
21\,cm fields can be made, as in \citet{ma_etal2018}. Instead, we opt to use the
Limber approximation in Equation~(\ref{eqn:limber}), which permits use of the
full 3D information present in the field. This approach produces bispectrum
values with sufficiently high S/N, as well as understanding the contribution
from each redshift to the total integrated signal.

Figure~\ref{fig:maps} shows the kSZ map and a slice through the 21\,cm field at
$z\sim 8$ subtending comparable solid angles of the sky. The large, degree-scale
differences between hot and cold patches in the kSZ map are related to the
large-scale velocity variations. The change to the kSZ signal induced by patchy
reionization from the EoR is evident on arcminute-scale features in the map,
comparable in shape to the neutral and ionized regions in the 21\,cm map.  The
expression of the bispectrum in Equation~(\ref{eqn:limber}) in the
squeezed-triangle limit relates the relatively large-scale modes in the 21\,cm
map with the small-scale ones in the kSZ map.

As a point of comparison, Figure~\ref{fig:maps} also shows the kSZ map filtered
such that only features on the scale of $\ell \sim 3000$ are preserved. This is
accomplished by convolving the kSZ map in $\ell$-space with a Gaussian window
centered about $\ell = 3000$ and a standard deviation of $\sigma_\ell = 100$. A
similar visualization of the 21\,cm field is generated for modes $\ell \sim 90$,
where the Gaussian window is centered about $\ell = 90$ with a standard
deviation of $\sigma_\ell = 30$. The statistic presented below quantifies the
degree of correlation between the large-scale modes of the 21\,cm map at
$\ell \sim 90$ and the small-scale power spectrum of the kSZ map at
$\ell \sim 3000$. We explore this correlation using a qualitative description of
reionization in Sec.~\ref{sec:toy_model}.

\subsection{Bispectrum Estimation}
\label{sec:bspec_calc}

As discussed in Sec.~\ref{sec:maps}, we opt to use the Limber approximation on
the right-hand side in Equation~(\ref{eqn:limber}) to compute the bispectrum as
opposed to directly evaluating the left-hand side. To calculate the bispectrum,
we use the so-called ``FFT-bispectrum'' estimator outlined in \citet{jeong2010}
and \citet{watkinson_etal2017}. This approach avoids explicitly enumerating all
triangles that contribute to a particular bisepctrum mode
$B(k_1,k_2,k_3)$. Instead, once the initial FFT is performed to yield a field
$\delta(\vb{k})$, the modes corresponding to $\delta(k_1)$, $\delta(k_2)$, and
$\delta(k_3)$ are stored in separate auxiliary fields in memory. Additional
``normalization fields'' with unit weight at these Fourier modes are also
generated for calculating the number of bispectrum triangles. An inverse FFT is
applied to each of these fields, and the cumulative sum of the product of the
cells in real space is computed, normalized by the sum of the product of the
normalization fields. The overall computational cost is significantly reduced
compared to explicit enumeration, and more accurate than Monte Carlo methods of
generating random triangle configurations.

Because the approach is based on repeated applications of the FFT rather than
enumerating triangle combinations, it can leverage computationally expedient and
highly optimized numerical libraries. \citet{watkinson_etal2017} demonstrated
good agreement between this estimator and the ``brute-force'' method of explicit
triangle enumeration. As an additional cross-check, we show validation results
of the FFT-bispectrum using the matter density field from an $N$-body simulation
and second-order cosmological perturbation theory in
Appendix~\ref{appendix:bspec}.

Evaluating Equation~(\ref{eqn:limber}) requires computing the 3D bispectrum at
fixed multipole moment $\ell=k \cdot \chi(z)$ as a function of
redshift. Accordingly, the modes $k_1$, $k_2$, $k_3$ change with co-moving
distance $\chi(z)$:
\begin{equation}
\chi(z) = c \int_0^z \frac{\dd{z}'}{H(z')},
\end{equation}
where $H(z)$ is the Hubble parameter. For the following analysis, we choose
$\ell_1 = 90$ and $\ell_2 = 3000$, and compute the bispectrum at 20 values of
$\theta_{12}$ evenly spaced in $\cos \theta$ between $-1$ and $1$. As mentioned
in Sec.~\ref{sec:bspec_background}, these $\ell$-scales correspond to those that
will be probed in upcoming experiments.

\section{Results}
\label{sec:results}

\subsection{Bispectrum Components}
\label{sec:full_bspec}

\begin{figure}
  \centering
  \includegraphics[width=0.45\textwidth]{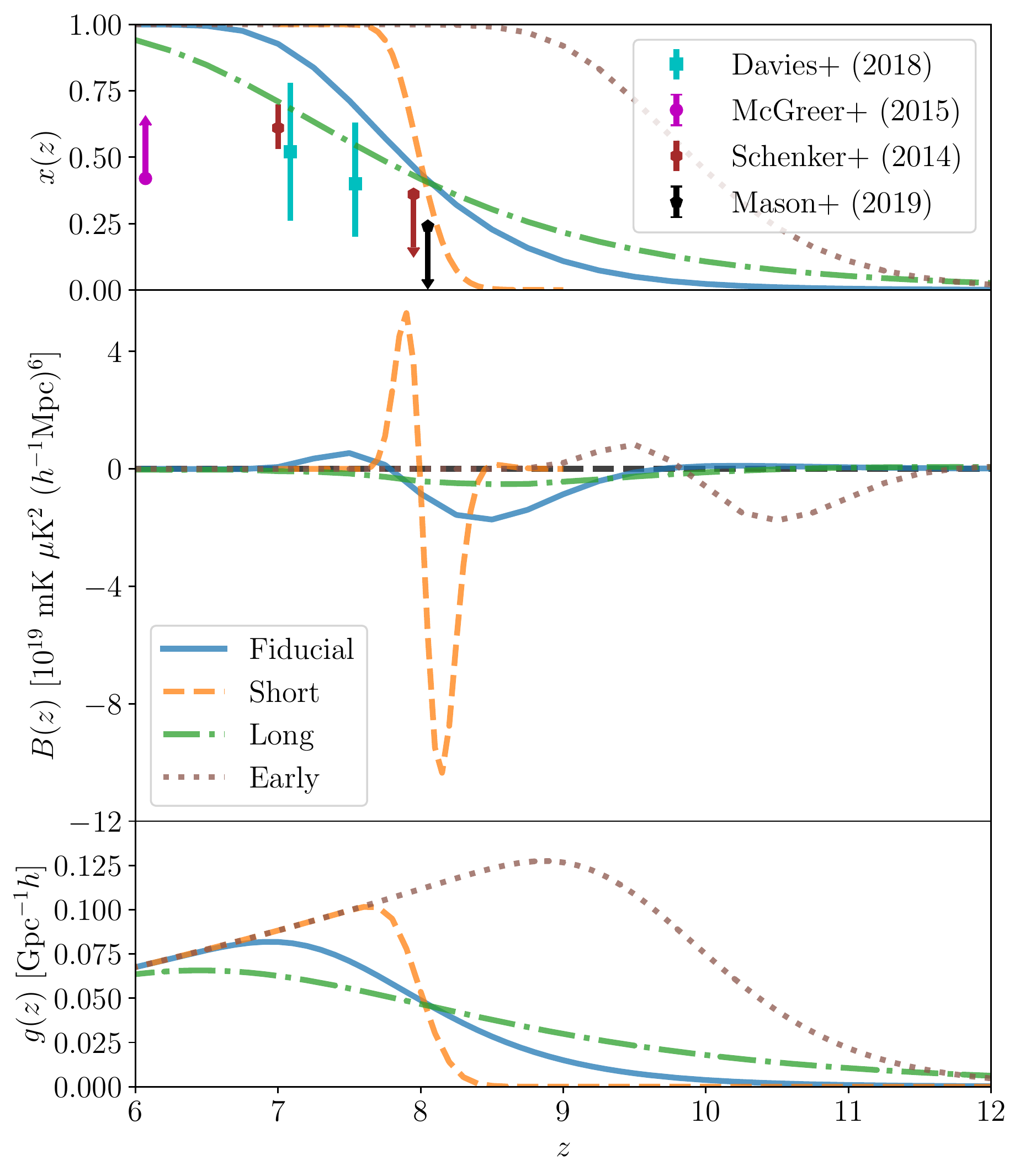}
  \caption{Top: global ionization fraction $x_i$ as a function of redshift $z$
    for the fiducial reionization history discussed in Sec.~\ref{sec:zreion}. In
    addition to the ionization histories for our simulations, we show
    observational constraints from various sources.  Center: the angle-averaged
    bispectrum $B_{\mathrm{21cm},q,q}$ for different ionization histories. We
    use the 3D analog of the kSZ field defined in Equation~(\ref{eqn:qksz}). The
    bispectrum is computed for $\ell_1 = 90$ and $\ell_2 = 3000$. Bottom: the
    kSZ window function defined in Equation~(\ref{eqn:wksz}). The bispectrum and
    window function, along with the 21\,cm window function, define the integrand
    in the Limber approximation in Equation~(\ref{eqn:limber}).}
  \label{fig:b_of_z}
\end{figure}

As shown in Equation~(\ref{eqn:limber}), the quantities that contribute to the
observable bispectrum
$\mathcal{B}_\mathrm{21cm,kSZ,kSZ}(\ell_1, \ell_2, \ell_3)$ are the 21\,cm
window function $W_\mathrm{21cm}(\chi)$, the kSZ visibility function $g(\chi)$,
the RMS velocity $v_\mathrm{RMS}^2$, and the 3D bispectrum
$B_{\mathrm{21cm},q,q}(k_1, k_2, k_3)$. The visibility function $g(\chi)$ is set
by the reionization history, and generally peaks near the end of
reionization. The RMS velocity can be computed from the peculiar velocities of
dark matter particles in the $N$-body simulation, and monotonically increases
over the redshifts considered here. The 21\,cm window function in practice is
defined by the observational strategy of the 21\,cm experiment and depends on
how data from different frequencies are combined. We discuss the window function
more in Sec.~\ref{sec:2dbspec}.

Figure~\ref{fig:b_of_z} shows the result of computing the 3D bispectrum
$B_{\mathrm{21cm},q,q}$ between the 21\,cm field $\delta T$ from
Equation~(\ref{eqn:t21}) and the 3D analog of the kSZ field $\delta_q$ defined
in Equation~(\ref{eqn:qksz}). The top panel shows the globally averaged
ionization fraction $x(z)$ for the volume. In addition to the histories from our
simulations, we show observational constraints on this quantity from various
sources. These are: IGM damping wing signatures from two quasars
\citep{davies_etal2018}; dark pixels in the Ly$\alpha$ and Ly$\beta$ forests
\citep{mcgreer_etal2015}; measurements of the fraction of Lyman-break selected
galaxies that emit prominent Ly$\alpha$ lines \citep{schenker_etal2014}; and the
Ly$\alpha$ equivalent width distribution \citep{mason_etal2019}. There are also
constraints on the optical depth to the CMB $\tau$ from Planck, which have been
further constrained by the EDGES high-band data \citep{monsalve_etal2019}. In
general, these measurements favor a relatively late end to reionization and a
relatively extended reionization history. While these are not physically
parameterized models, or necessarily chosen to be consistent with current
observational constraints, the important point is that they produce internally
consistent 21\,cm brightness temperature and kSZ observables, and that they show
the qualitative dependence of the bispectrum signal on the duration and timing
of reionization. The fiducial case is chosen to produce a plausible amplitude
(which is enhanced for fast scenarios and suppressed for slow ones), though the
timing most consistent with data would be somewhat later.  Thus, the
21\,cm--kSZ--kSZ bispectrum can be a valuable approach that can help corroborate
our current understanding of the reionization history. The middle panel is the
quantity $B_{\mathrm{21cm},q,q}$ from Equation~(\ref{eqn:limber}), with
$\ell_1 = 90$ and $\ell_2 = 3000$. The plotted quantity is the angle-averaged
bispectrum, which is a weighted average over all angles $\theta_{12}$ defined in
Equation~(\ref{eqn:theta}). The bottom panel is the kSZ window function
$g(\chi)$ defined in Equation~(\ref{eqn:wksz}) for each reionization history.

The amplitude of the bispectrum reaches a maximum shortly before the midpoint of
reionization at $\ev{x_i} \sim 0.25$, after which point the amplitude
decreases. The sign of the bispectrum is negative, signifying that the fields
are highly anticorrelated. This result makes some intuitive sense: the 21\,cm
signal comes from neutral regions of the IGM, and the high amplitude of the kSZ
power spectrum on small scales comes from ionized ones. Thus, portions of the
IGM that have an above-average 21\,cm signal on large scales (i.e.,
$\delta T_b(k_1) > 0$) have a below-average contribution to the kSZ signals on
small scales. Because these regions are more neutral than average on large
scales, it follows that there are fewer highly ionized regions on smaller
scales, due to the inside-out reionization scenario implied by the semi-analytic
model. Also worth noting is that the amplitude of the signal depends strongly on
the duration of reionization, with shorter reionization scenarios yielding a
larger magnitude. In the semi-analytic model used here, shorter reionization
histories feature larger regions of neutral and ionized gas, which amplifies the
anticorrelation implied by the bispectrum. Conversely, the timing of
reionization does not significantly affect the amplitude of the bispectrum: the
fiducial and early histories have nearly identical shapes, and are just offset
in redshift. This feature implies that the amplitude of the integrated
bispectrum is largely driven by the duration of reionization, and the timing
affects which redshift windows would be sensitive to the bispectrum signal.

Interestingly, following the midpoint of reionization, the bispectrum
transitions sign, signifying that the fields are positively correlated. The
amplitude of this peak reaches a maximum at $\ev{x_i} \sim 0.75$, though its
amplitude is smaller than the anticorrelation near the midpoint of
reionization. As with the amplitude of the anticorrelation peak near the
midpoint of reionization, the amplitude of the positively correlated peak is
larger for shorter reionization histories. Notably, the positive correlation is
largely absent from the long reionization history, suggesting that for a gradual
enough reionization process, the bispectrum demonstrates only a (weakly)
anticorrelated signal. Further exploration of the trends seen in the behavior
of the bispectrum are explored in Sec.~\ref{sec:toy_model}.

The bottom panel of Figure~\ref{fig:b_of_z} shows the kSZ visibility function
$g(\chi)$. This quantity peaks near the earliest redshift associated with total
ionization of $\ev{x_i} \sim 1$. As can be seen by the functional form in
Equation~(\ref{eqn:wksz}), its value depends linearly on the global ionization
fraction, meaning that higher ionization fractions contribute more to the total
integral. At the same time, the window function depends quadratically on
redshift, and so once the universe is nearly totally ionized, the amplitude of
the window function decreases. For the fiducial ionization history, the
visibility function peaks near $z \sim 7$.

Additionally, the maximum value of the visibility function is larger for
histories that reach total ionization at earlier redshifts: the maximum value
for the early history is roughly a factor of two larger than that of the late
history. Again, due to the quadratic dependence of $g(\chi)$ on redshift,
earlier reionization histories result in a larger value for this
quantity. Earlier reionization histories also imply larger values of $\tau$,
which is simply the integral of the visibility function. At the same time, the
factor of $\chi^4$ in the denominator of the integrand of
Equation~(\ref{eqn:limber}) means that lower-redshift contributions are weighted
significantly more than higher-redshift ones, and so the higher-redshift
contributions are not necessarily weighted as significantly as the visibility
function itself might suggest.

\begin{figure}[t]
  \centering
  \includegraphics[width=0.45\textwidth]{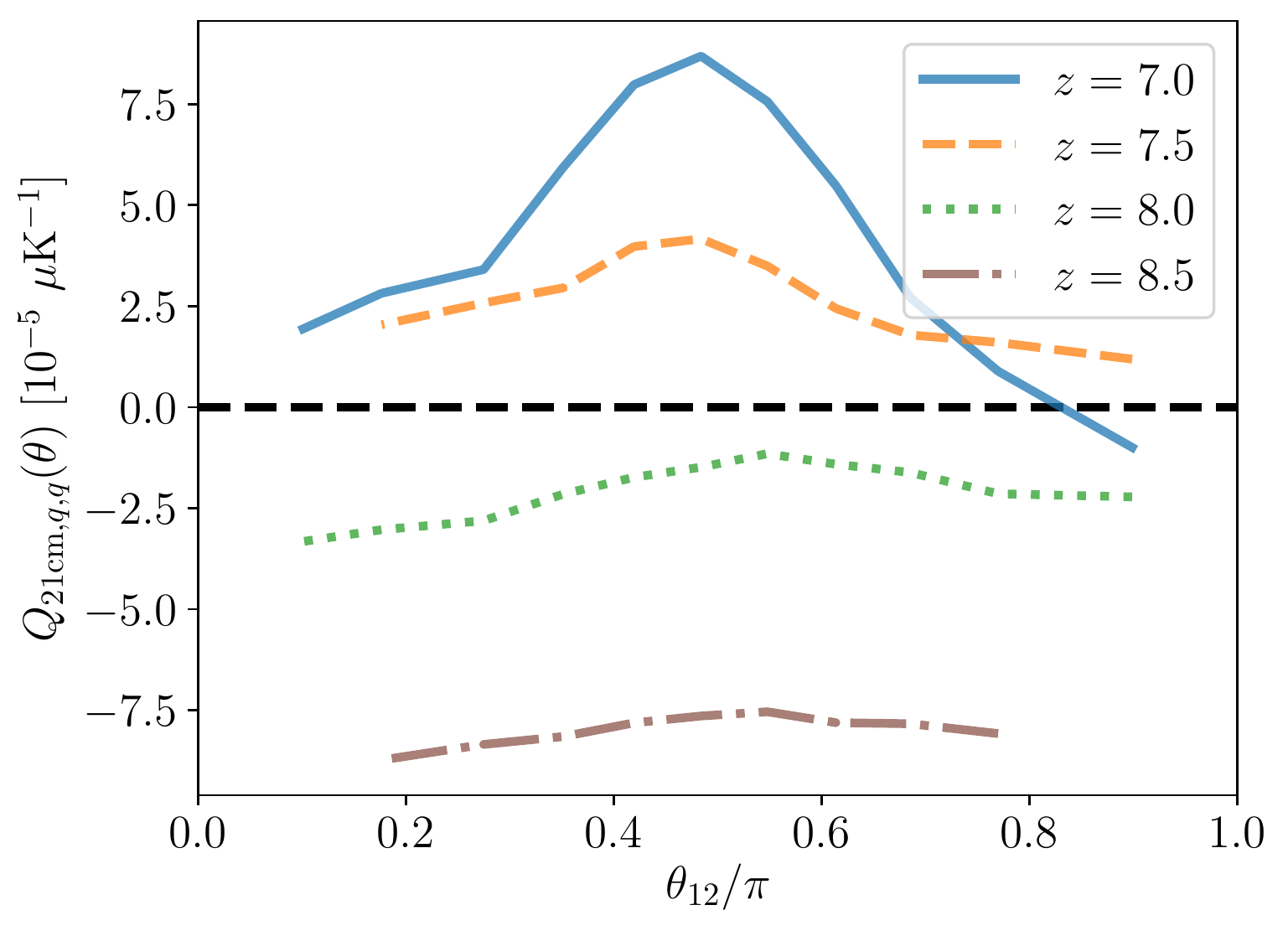}
  \caption{The 21\,cm $q$-$q$ bispectrum shown in Figure~\ref{fig:b_of_z},
    plotted as a function of angle $\theta_{12}$ defined in
    Equation~(\ref{eqn:theta}). There is significant evolution of the bispectrum
    as a function of angle $\theta_{12}$. At early times ($\ev{x_i} < 0.5$), the
    amplitude of the bispectrum as a function of $\theta_{12}$ is nearly
    constant. However, at late times ($\ev{x_i} > 0.5$) the amplitude of the
    bispectrum is significantly greater when $\theta_{12} \sim \pi/2$. These
    correspond to configurations where the triangles are roughly isosceles,
    which are more sensitive to correlations in void regions as opposed to
    filamentary structure. This behavior is different than the matter bispectrum
    shown in Appendix~\ref{appendix:bspec}. For additional discussion, see
    Sec.~\ref{sec:full_bspec}.}
  \label{fig:b_of_theta}
\end{figure}

To understand how the magnitude of the bispectrum changes as a function of angle
$\theta_{12}$, we compute the reduced bispectrum $Q_{abc}(k_1, k_2, k_3)$,
defined as:
\begin{multline}
Q_{abc}(k_1, k_2, k_3) \equiv \\
\frac{B_{abc}(k_1, k_2, k_3)}{P_{aa}(k_1)P_{bb}(k_2) + P_{bb}(k_2)P_{cc}(k_3) + P_{aa}(k_1)P_{cc}(k_3)}
\label{eqn:reduced_bspec}
\end{multline}
where $a,b,c$ are different fields of interest, $B_{abc}(k_1,k_2,k_3)$ is the
bispectrum, and $P_{aa}(k_1)$ is the value of the auto-power spectrum of the
field $a$ at a value of $k_1$. The reduced bispectrum removes some of the large
amplitude differences present at different scales, especially for
squeezed-triangle configurations. In the case of the bispectrum under
consideration here, it also allows for straightforward comparison of the signal
between different redshifts, where the magnitude of $B$ is very different. To
allow for a more even comparison, we multiply the bispectrum
$B_{\mathrm{21cm},q,q}$ by $v_\mathrm{RMS}^2/3c^2$, and the power spectrum
$P_{qq}$ by the same factor. Also, we plot the result of computing
$B(k_1, k_2, \theta_{12})$, using $\theta_{12}$ as defined in
Equation~(\ref{eqn:theta}). As in the above analysis, we use $k_1 = \ell_1/\chi$
and $k_2 = \ell_2/\chi$, where $\ell_1 = 90$, $\ell_2 = 3000$, and $\chi$ is the
co-moving distance to redshift $z$. Note that the definition of $Q$ in
Equation~(\ref{eqn:reduced_bspec}) yields a quantity that has units of inverse
temperature for bispectra and power spectra of fields that have temperature
units. We convert all temperature units to $\mu$K before combining quantities,
and show overall results in units of $(\mu \mathrm{K})^{-1}$.

\begin{figure*}[t]
  \centering
  \includegraphics[width=0.32\textwidth]{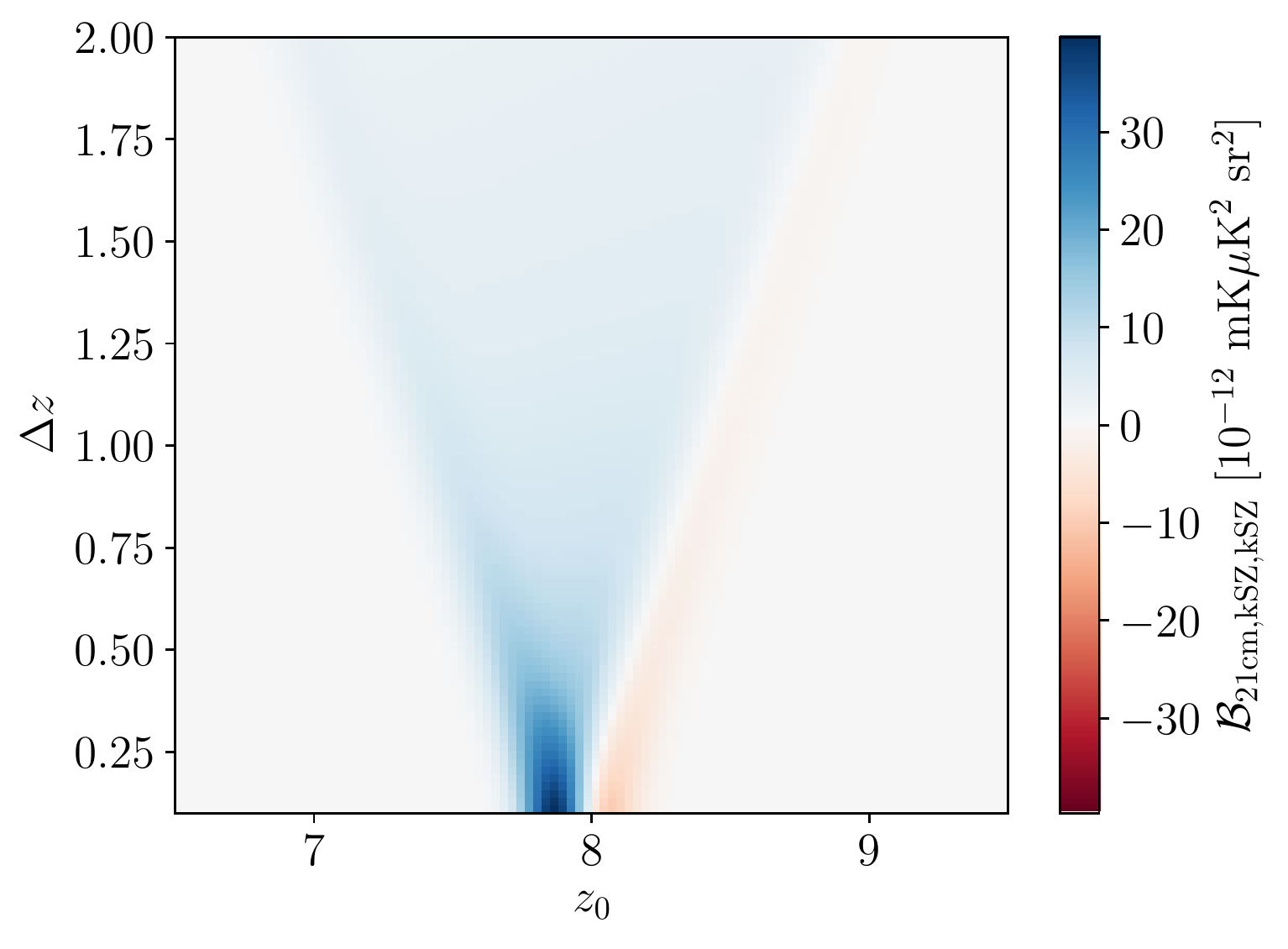} \hfill
  \includegraphics[width=0.32\textwidth]{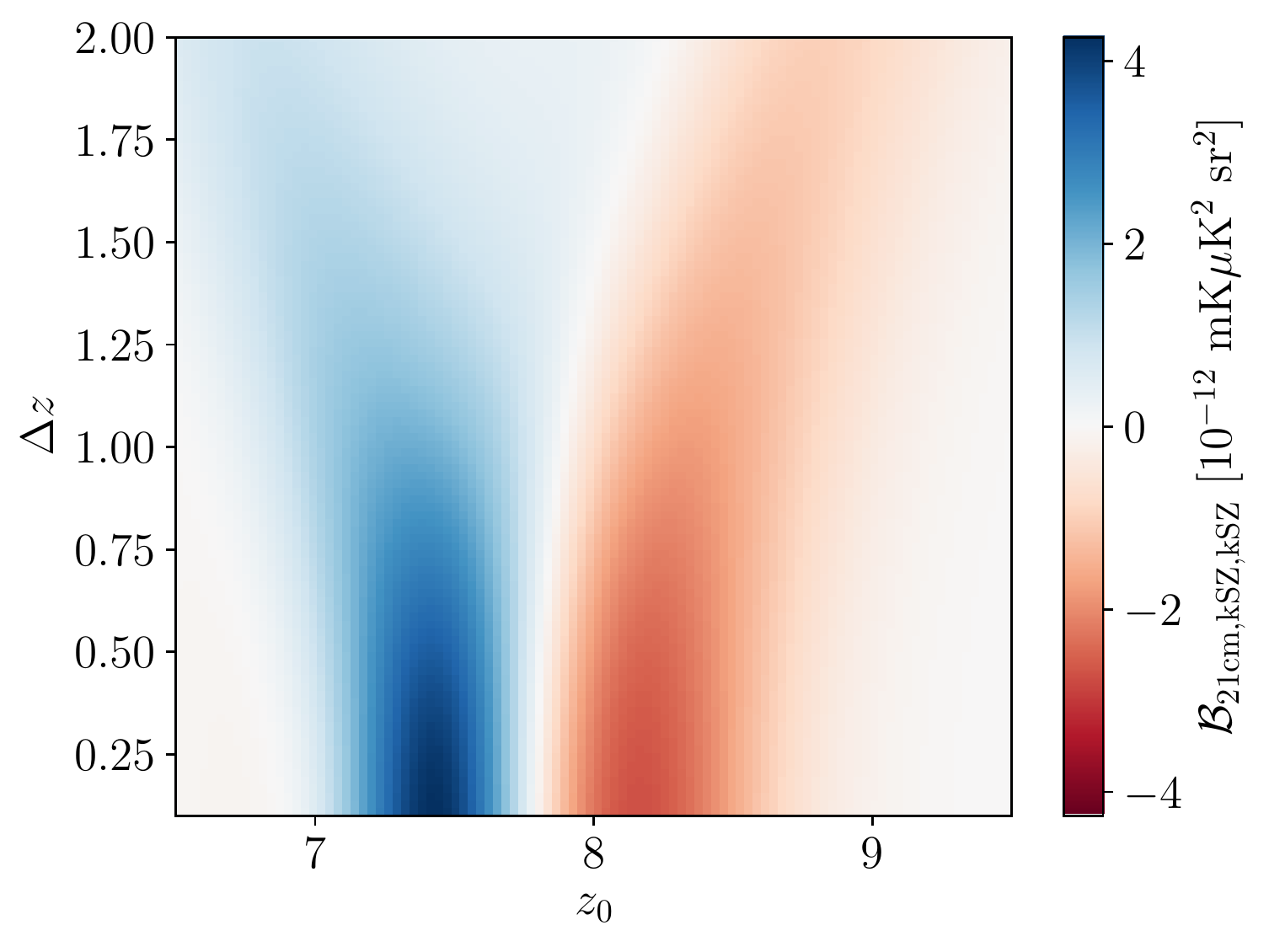} \hfill
  \includegraphics[width=0.32\textwidth]{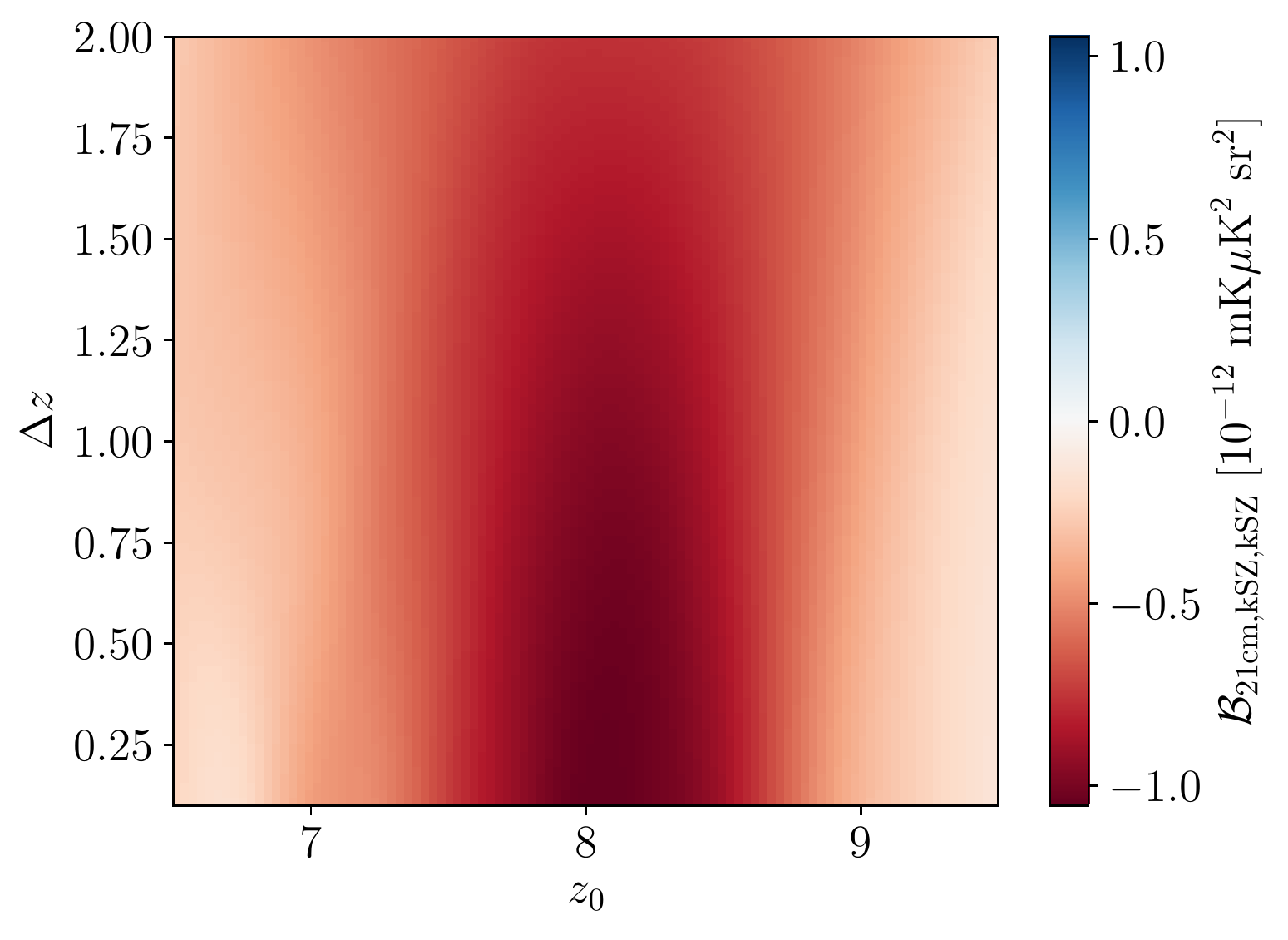}
  \caption{The result of computing the Limber integral defined in
    Equation~(\ref{eqn:limber}) as a function of the 21\,cm window function
    $W_\mathrm{21cm}(\chi)$ defined in Equation~(\ref{eqn:w21cm}). We
    parameterize the 21\,cm window function as a simple top-hat in co-moving
    distance $\chi$, with a central redshift of $z_0$ and a total width of
    $\Delta z$. Left: the short reionization history. Center: the fiducial
    reionization history. Right: the long reionization history. The ionization
    history, 3D bispectrum, and kSZ visibility functions can be seen in
    Figure~\ref{fig:b_of_z}. The duration of the reionization history has
    interesting implications for the amplitude and behavior of the
    statistic. See Sec.~\ref{sec:2dbspec} for full discussion.}
  \label{fig:bspec_limber}
\end{figure*}

Figure~\ref{fig:b_of_theta} shows the results of computing $Q(\theta)$ according
to Equation~(\ref{eqn:reduced_bspec}) for the 21\,cm field and proxy for the kSZ
field defined in Equation~(\ref{eqn:qksz}). We have plotted the quantity for the
fiducial reionization history at the indicated redshift values, which includes
the majority of the reionization history. At the earliest redshift $z = 8.5$
corresponding to $\ev{x_i} \sim 0.1$, there is little dependence of this
quantity on angle $\theta_{12}$. However, at later times, the quantity shows
significant differences as a function of $\theta_{12}$. When the bispectrum is
negative (corresponding to times prior to the midpoint of reionization), the
more extreme values correspond to angles $\theta \sim 0$ or $\theta \sim
\pi$. Triangles with these values are more sensitive to structure that lies
along filaments, meaning that the signal is likely more sensitive to the kSZ
signal that lies along these overdensities. When the bispectrum is positive
(post-reionization), the more extreme values are seen for angles of
$\theta \sim \pi/2$. These triangles are more sensitive to isotropic
distributions, and are likely responding to the relatively large ionized regions
that the 21\,cm mode $k_1$ is sensitive to.

\subsection{The Integrated Bispectrum}
\label{sec:2dbspec}
As mentioned in Sec.~\ref{sec:bspec_background}, the Limber approximation in
Equation~(\ref{eqn:limber}) can be used to convert the 3d 21\,cm--kSZ--kSZ
bispectrum shown in Figure~\ref{fig:b_of_z} to a 2D version that would be
suitable for comparing with measurements from upcoming surveys. Also as
discussed above, computing the integral as defined in
Equation~(\ref{eqn:limber}) requires defining the 21\,cm window function
$W_\mathrm{21cm}(\chi)$. The only formal requirement that is imposed is that the
window function integrates to unity: $\int \dd{\chi} W_\mathrm{21cm}(\chi) = 1$.
In principle, one can choose the window function to maximize the theoretical
response of the full Limber integral in Equation~(\ref{eqn:limber}) given the 3D
bispectrum $B_{\mathrm{21cm},q,q}$ and the kSZ visibility function $g(\chi)$. In
practice, we use a simple top-hat window function for $W_\mathrm{21cm}(\chi)$ in
the following analysis to understand how the full result depends on the choice
of the window function. This approach is computationally simpler, though it does
not combine the 21\,cm signal at multiple redshifts/frequencies with equal
weightings. For more fidelity with upcoming surveys, one could instead choose a
window function analogous to using, e.g., a Blackman-Harris window
function. However, we leave such considerations to future analysis.

We parameterize the window function $W_\mathrm{21cm}(\chi)$ as a top-hat
centered on a redshift $z_0$ with a total width of $\Delta z$. Mathematically,
this can be expressed as:
\begin{align}
W_\mathrm{21cm}(z) &= W_0 \Theta\qty(z - (z_0 - \Delta z / 2)) \notag \\
& \qquad \times \Theta\qty((z_0 + \Delta z / 2) - z),
\label{eqn:w21cm}
\end{align}
where $\Theta(z)$ is the heaviside theta function. The normalization $W_0$ has
units of inverse length (e.g., Mpc$^{-1}h$) such that the integral in
co-moving distance $\chi$ is unity. In the following figures, we show results
related to integrating Equation~(\ref{eqn:limber}) with different choices of
$z_0$ and $\Delta z$.

One caveat with using the Limber approximation for computing the $C_\ell$
spectrum is that the total window of integration should be much larger than the
target $k$-mode chosen (i.e., $\Delta \chi \gg 1/k$). For $\ell_1 = 90$, this
corresponds to $k$-modes of magnitude $k \sim 0.01$ Mpc$^{-1}h$. Accordingly,
the window of integration should be $\Delta z \gtrsim 0.5$ for the redshifts of
interest here. As explained more in Sec.~\ref{sec:snr_noise_free}, the $C_\ell$
spectrum is only relevant when forecasting the signal-to-noise ratio of the
bispectrum. The bispectrum signal itself is not subject to the same restriction,
due to the presence of only a single 21\,cm field. Accordingly, in the figures
below, we include windows down to $\Delta z = 0.1$. At the same time, we have
noticed that computing the $C_\ell$ spectrum of the 21\,cm field from the
two-point Limber approximation (similar to the one in
Equation~(\ref{eqn:clksz})) shows qualitatively good agreement for most $\ell$
modes.

Figure~\ref{fig:bspec_limber} shows the expected signal from integrating
Equation~(\ref{eqn:limber}) with different choices of the 21\,cm window function
$W_\mathrm{21cm}(\chi)$ as defined in Equation~(\ref{eqn:w21cm}). The duration
of reionization strongly impacts the amplitude of the cross-bispectrum
$\mathcal{B}_\mathrm{21cm,kSZ,kSZ}$. For the short history, the amplitude of the
resulting signal is the largest compared to the other histories, and is seen
almost exclusively as a positive correlation. The signal is maximized for a
relatively narrow integration window $\Delta z$, and has a maximal response when
centered shortly after the midpoint of reionization. This result makes sense
given the functional forms of the 3d bispectrum $B_{\mathrm{21cm},q,q}$ and
visibility function $g(\chi)$ seen in Figure~\ref{fig:b_of_z}. Given the steep
increase in $g(\chi)$ as a function of $z$, the statistic is most sensitive to
the positive correlation following the midpoint of reionization as opposed to
the anticorrelation prior to the midpoint. The short history also has the
largest amplitude of the underlying 3D bispectrum $B$, and so has the largest
amplitude in the integrated bispectrum $\mathcal{B}$. The statistic also has a
larger amplitude for narrower windows, which permits the inclusion of primarily
the positive correlation following the midpoint of reionization while excluding
the anticorrelation prior to the midpoint. A wide window in redshift includes
contributions from both, reducing the amplitude of the resulting signal. Thus,
narrower observational windows in redshift yield a larger signal.

In the fiducial reionization history, the integrated bispectrum $\mathcal{B}$
exhibits interesting behavior as a function of the center of integration
$z_0$. When the window is centered on redshifts corresponding to redshifts prior
to the midpoint of reionization, the resulting statistic is negative. This
coincides with the large, negative value of the 3D bispectrum $B$ at these
redshifts. As the center of the window $z_0$ shifts to later times, the value of
$\mathcal{B}$ becomes positive. Similar to the result of the short reionization
history, windows centered on these times receive the most contribution from the
positive correlation following the midpoint of reionization as opposed to the
anticorrelation prior to the midpoint. As with the short history, narrower
windows in redshift yield more extreme values in the resulting statistic, again
arguing for narrow observing windows to maximize the resulting signal. However,
the overall amplitude of the signal is smaller for the fiducial history compared
to the short history, largely due to the smaller amplitude of the 3D bispectrum
$B$.

In the long reionization history, the integrated bispectrum $\mathcal{B}$ is
negative for all choices of the 21\,cm window function. This is a result of the
3D bispectrum $B$ being negative for all redshifts. In this reionization
history, the 21\,cm and kSZ fields are always anticorrelated, and do not
demonstrate a positive correlation. Additionally, the maximal response occurs
when the window is centered near the midpoint of reionization, rather than
significantly before or after. The amplitude of the statistic is also smaller
than that of the other reionization history, though it falls off less
significantly with the width of the window function $\Delta z$.

An interesting result of the varying reionization history is that the behavior
of the integrated bispectrum $\mathcal{B}$ is very sensitive to the duration of
reionization. For relatively short histories, the bispectrum $\mathcal{B}$ has
positive amplitude and exhibits the largest response, but is quite sensitive to
both the central value and width of the 21\,cm window function. For moderate
duration histories such as those in our fiducial history, the sign of the
bispectrum $\mathcal{B}$ changes depending on the center of the 21\,cm window
function. The amplitude is comparable between the two cases, though larger for
the post-reionization window. For relatively long reionization histories, the
integrated bispectrum $\mathcal{B}$ does not change sign, and is negative
regardless of the parameters of the 21\,cm window function. As we explore in
greater detail below, much of this behavior can be attributed to the degree of
correlation between the 21\,cm field on large scales and the density and
ionization fields on small scales. In general, for longer histories, the degree
of correlation is weaker, and so the magnitude of the signal is smaller than for
short reionization histories.

\subsection{Bispectra of Ionization and Density Fields}
\label{sec:linear_bspec}

The full kSZ field requires computing the product of the ionization field and
matter field in real space before applying an FFT as in
Equation~(\ref{eqn:ft}). However, some intuition can be gleaned from examining
the behavior of the bispectrum between the 21\,cm field and the fields
contributing to the kSZ effect (Equation~\ref{eqn:qksz}). By understanding how
various combinations of the ionization and matter density fields evolve as a
function of redshift, we can explain some features in the full bispectrum seen
in Figures~\ref{fig:b_of_z} and \ref{fig:b_of_theta}. To probe this, we compute
the bispectrum between various permutations of the 21\,cm field at $\ell_1 = 90$
and the ionization or matter fields at $\ell = 3000$. Equation~(\ref{eqn:ksz})
shows how the kSZ field can be expressed as the product of the ionization field
and density contrast, and so some of the behavior of the full bispectrum can be
explained by the behavior of the correlation between the 21\,cm field and
combinations of the ionization field $x_i$ and matter density field
$\delta_m$. For instance, we compute the bispectrum in the squeezed-triangle
limit of the 21\,cm field and two fields of the local ionization $x_i$, denoted
as $B_{\mathrm{21cm},x,x}$:
\begin{multline}
\ev{\tilde{\delta T}_\mathrm{21cm}(\vb{k}_1) \tilde{x}_i(\vb{k}_2) \tilde{x}_i(\vb{k}_3)} \equiv \\
(2\pi)^3 \delta_D(\vb{k}_1 + \vb{k}_2 + \vb{k}_3) B_{\mathrm{21cm},x,x}(k_1,k_2,k_3).
\end{multline}
By analogy, we also compute the bispectrum involving two fields of the matter
overdensity $\delta_m$ which we denote as $B_{\mathrm{21cm},\rho,\rho}$. We also
compute the cross-spectra between the ionization field and the matter density
field $B_{\mathrm{21cm},x,\rho}$ and $B_{\mathrm{21cm},\rho,x}$.  Note that the
definition of the bispectrum does not lead to equality when interchanging the
indices, i.e., $B_{\mathrm{21cm},x,\rho} \neq B_{\mathrm{21cm},\rho,x}$. These
quantities are essentially the same when computing an angle-averaged quantity,
and have opposite behavior as a function of $\theta_{12}$. (For instance, if
$B_{\mathrm{21cm},x,\rho}$ increases as a function of $\theta_{12}$, then
$B_{\mathrm{21cm},\rho,x}$ decreases by an equal and opposite amount as a
function of $\theta_{12}$.) For the sake of brevity in the results below, we
only show the quantity $B_{\mathrm{21cm},x,\rho}$.

\begin{figure}[t]
  \centering
  \includegraphics[width=0.45\textwidth]{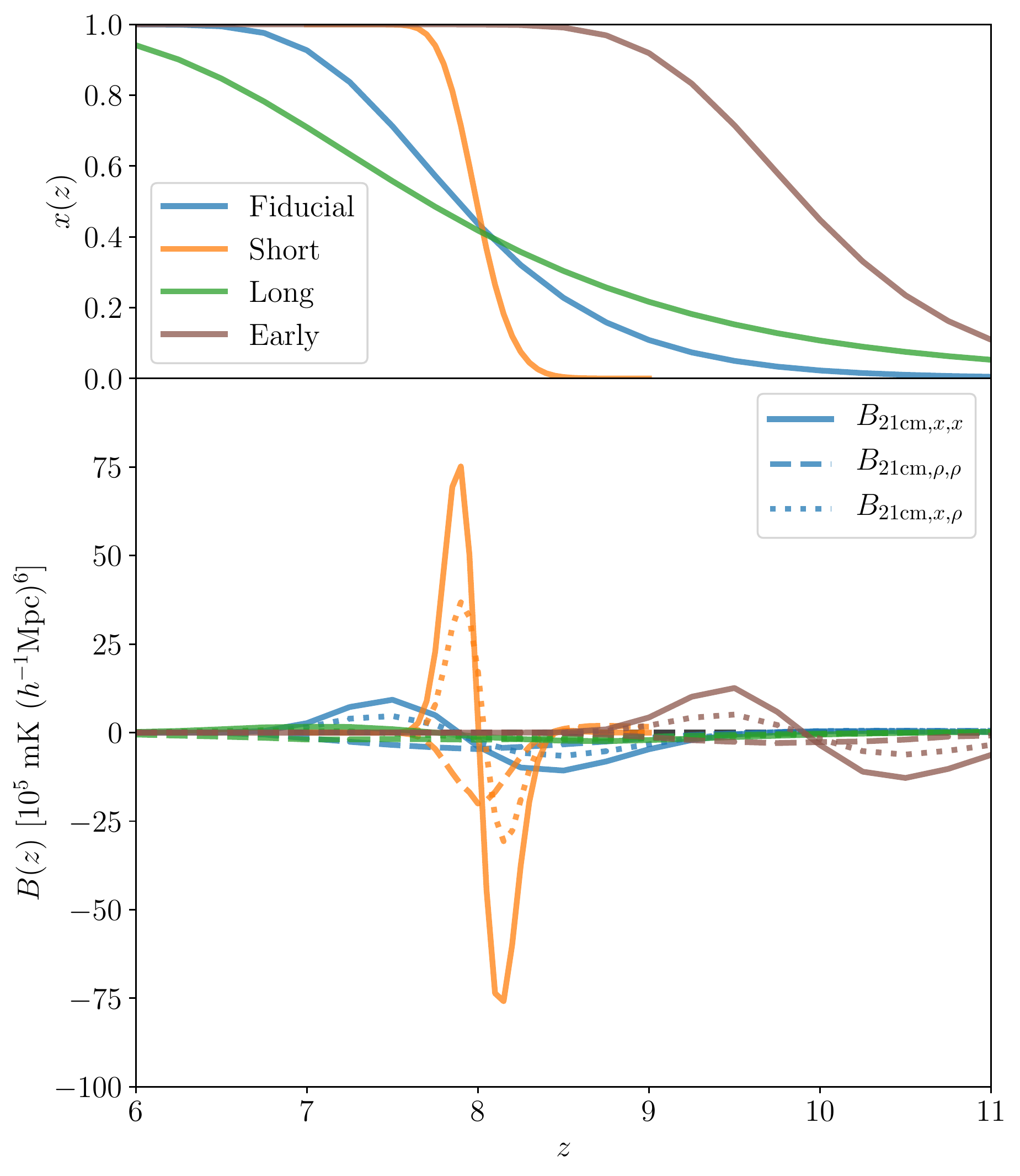}
  \caption{Top: the global ionization fraction $x(z)$. Bottom: the
    angle-averaged squeezed-triangle bispectrum between the 21\,cm field and
    combinations of the ionization field $x_i$ and the matter overdensity
    $\delta_m$. Different histories are plotted using different colors and
    different combinations of constituent fields are shown in different line
    styles. Bispectra involving the ionization field $x_i$ transition from
    negative to positive as ionization progresses, transitioning shortly after
    the midpoint. The bispectrum with only the matter field is negative at all
    times. This behavior of the bispectrum can be explained by a qualitative
    description of reionization discussed in Sec.~\ref{sec:toy_model}.}
  \label{fig:b_of_z_components}
\end{figure}

As mentioned in Sec.~\ref{sec:bspec_background}, we are computing the bispectrum
in the squeezed-triangle limit using the 21\,cm field as the short leg of the
triangle; accordingly, the bispectrum can be thought of as a conditional probe
of the small-scale structure given a particular value for the large-scale 21\,cm
field. Furthermore, because the squeezed-triangle configuration guarantees that
$k_2 \approx k_3$, this conditional probe of small-scale structure is
proportional to the amplitude of the power spectrum on the scale $k_2$. For
example, if $B_{\mathrm{21cm},x,x}(k_1, k_2, k_3)$ has a large positive
amplitude, this means that large-scale regions of an above-average 21\,cm
brightness temperature $\delta T(k_1)$ are correlated with a relatively
high-amplitude power spectrum of the ionization field $P_{xx}(k_2)$---at the
same time, large-scale regions with below-average 21\,cm brightness temperature
$T_\mathrm{21cm}(k_1)$ are correlated with a low-amplitude power spectrum
$P_{xx}(k_2)$. A negative amplitude implies an anticorrelation between the
large-scale amplitude of the 21\,cm field and the power spectrum of the
small-scale modes. If there is no statistical relationship between the
large-scale field amplitude and small-scale power spectrum amplitude, then the
bispectrum amplitude is near 0.

\begin{figure}[t]
  \centering
  \includegraphics[width=0.45\textwidth]{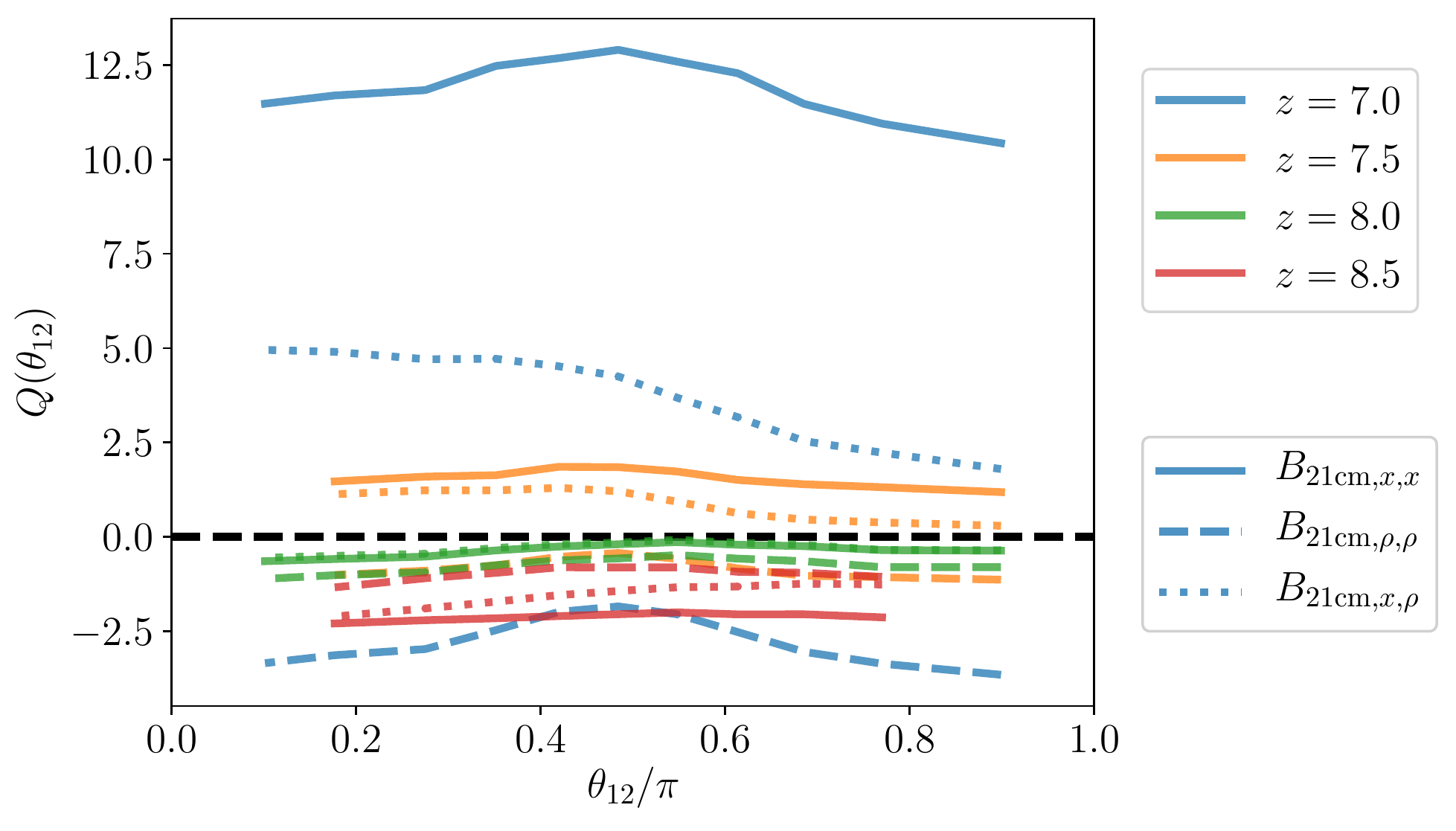}
  \caption{The reduced bispectrum (defined in
    Equation~(\ref{eqn:reduced_bspec})) for the components shown in
    Figure~\ref{fig:b_of_z_components} as a function of angle $\theta_{12}$
    defined in Equation~(\ref{eqn:theta}). Different combinations of constituent
    bispectra are shown in different line styles, with different colors
    representing different redshift values. All bispectra here are from the
    fiducial reionization history, though the other histories show qualitatively
    similar results at comparable ionization fractions. At early times, the
    bispectra show little sensitivity to the angle $\theta_{12}$. At later
    times, the bispectrum is more sensitive to the orientation of the triangles,
    though not nearly to the same degree as the integrated bispectrum shown in
    Figure~\ref{fig:b_of_theta}. This is primarily due to the fact that
    Figure~\ref{fig:b_of_theta} is the sum of the individual terms shown
    here. While each of these terms is slightly peaked near
    $\theta_{12} \sim \pi/2$, the sum of these terms yields a result that has a
    more prominent angular dependence.}
  \label{fig:b_of_theta_components}
\end{figure}

Figure~\ref{fig:b_of_z_components} shows the squeezed-triangle bispectrum
between the 21\,cm field and various combinations of the ionization field $x_i$
and matter contrast $\delta_m$. The quantity plotted is
$B(k_1, k_2, \theta_{12})$, with an average performed over all angles
$\theta_{12}$ weighted by the number of triangles that generate a particular
combination, similar to the central panel of Figure~\ref{fig:b_of_z}. In the
earliest stages of reionization, all of the component bispectra are negative,
meaning that the amplitude of the small-scale power spectra are anticorrelated
with the large-scale amplitude of the 21\,cm field. At the midpoint of
reionization, there is a rapid transition for the component bispectra that
includes the ionization field. Although these ionization bispectra become
positively correlated, $B_{\mathrm{21cm},\rho,\rho}$ remains negatively
correlated. Following the midpoint and continuing to the end of reionization,
combinations with the ionization field $x_i$ are positive, and ones with
exclusively the density field remain negative. This qualitative picture is true
for all reionization histories, though the amplitude is larger for shorter
histories. We develop a qualitative description to explain these features below
in Sec.~\ref{sec:toy_model}.

Figure~\ref{fig:b_of_theta_components} shows the component bispectra from
Figure~\ref{fig:b_of_z_components} but as a function of angle $\theta_{12}$. In
this figure, we also have transitioned to plotting the reduced bispectrum
$Q(\theta_{12})$ defined in Equation~(\ref{eqn:reduced_bspec}). To remove the
impact of temperature normalization, we have divided the bispectra by $T_0(z)$
defined in Equation~(\ref{eqn:t0}) and divided the power spectra
$P_\mathrm{21cm,21cm}$ by $T_0(z)^2$. In this case, the quantity
$Q(\theta_{12})$ is dimensionless instead of having temperature units as in
Sec.~\ref{sec:full_bspec}. At early times, $Q(\theta_{12})$ is relatively flat
as a function of $\theta_{12}$. This flatness means there is not significant
preferential alignment between the 21\,cm field and the ionization or density
fields. Following the midpoint of reionization, the bispectra begin to show
evolution with angle. At late times ($z = 7$), the bispectrum shows the most
significant evolution as a function of angle. The ionization-only bispectrum
$B_{\mathrm{21cm},x,x}$ has its largest response near isosceles triangles
($\theta_{12} \sim \pi/2$), whereas the density-only bispectrum
$B_{\mathrm{21cm},\rho,\rho}$ has the most extreme values for oblique triangles
($\theta_{12} \sim 0$ and $\theta_{12} \sim \pi$). Note that the sign is
different in these two cases, so that summing the individual bispectra
components yields a quantity that has a larger amplitude for isosceles compared
to oblique triangles. This helps explain why the angular dependence of the full
bispectrum (a sum of the terms shown above and others) is more extreme than
these individual components.

\subsection{Qualitative Behavior of the Bispectrum}
\label{sec:toy_model}

\begin{figure*}[t]
  \centering
  \includegraphics[width=0.9\textwidth]{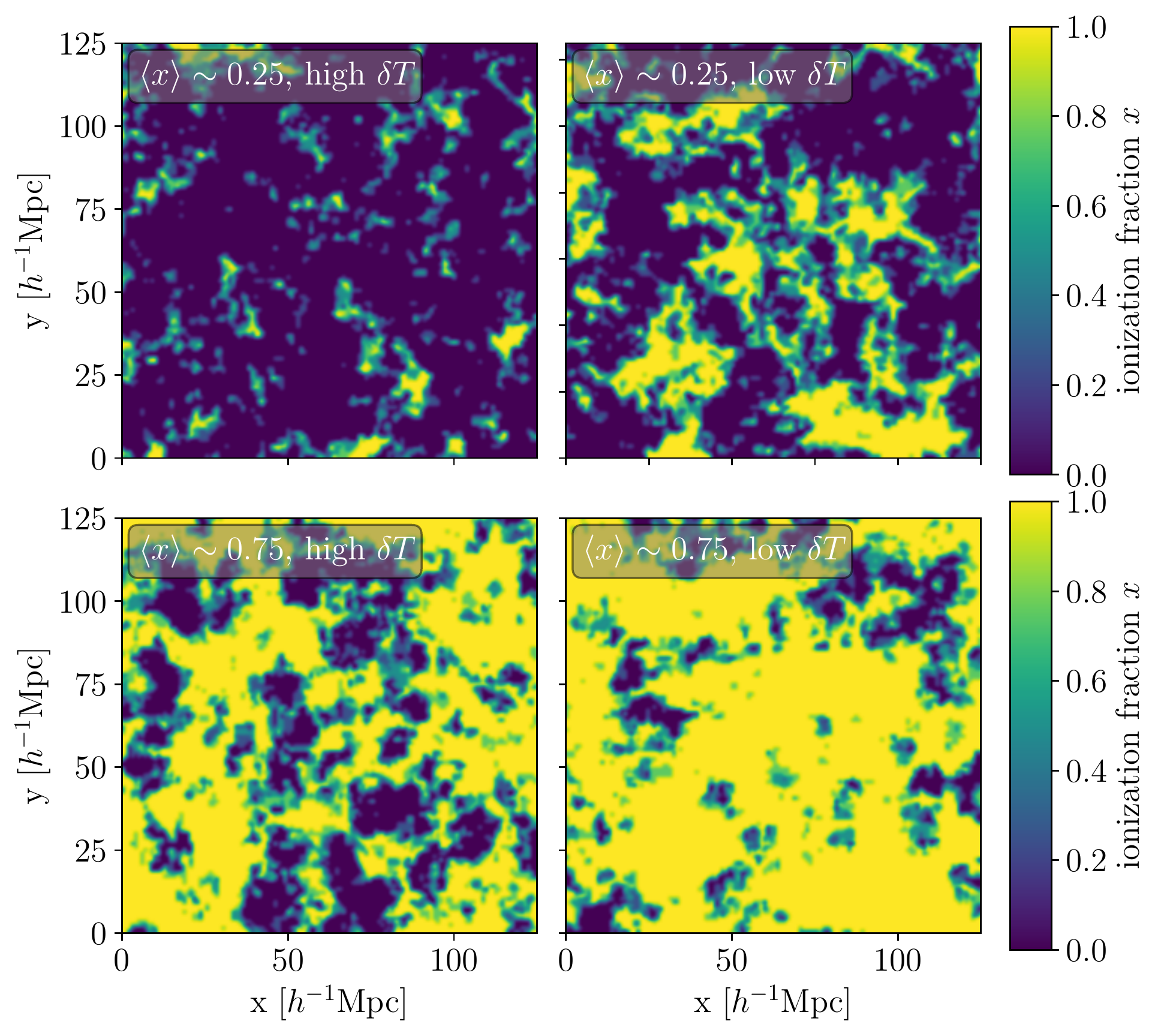}
  \caption{A visualization through the simulation volume at $z = 8.5$
    ($\ev{x_i} \sim 0.25$, top) and $z = 7.5$ ($\ev{x_i} \sim 0.75$,
    bottom). The left column shows a slice through a subregion of $L = 125$
    $h^{-1}$Mpc for which the local average 21\,cm spin temperature $\delta T_b$
    is above the global average for the full $L = 2$ $h^{-1}$Gpc volume, and the
    right column shows a slice through a subregion where the local average
    $\delta T_b$ is below the global average.  At early times, the regions of
    high $\delta T_b$ have fewer ionized regions and less small-scale ionization
    power. Therefore, $B_\mathrm{21cm,x,x}$ is initially negative. At late
    times, the regions of high $\delta T_b$ are less ionized than average, yet
    have more small-scale structure left over in the ionization field. The low
    $\delta T_b$ regions are mostly ionized, with less small-scale structure
    remaining in the ionization field. Correspondingly, $B_\mathrm{21cm,x,x}$ is
    positive at late stages of reionization.}
  \label{fig:tslices}
\end{figure*}

The results in the previous section can be understood with the help of a simple
qualitative description of reionization. We begin with an explanation of the
behavior of the constituent bispectra at early times before the midpoint of
reionization. As in Sec.~\ref{sec:bspec_background}, we take $k_1$ to be the
Fourier mode corresponding to the 21\,cm field, and it probes relatively large
scales ($k_1 \sim 0.05$ Mpc$^{1}~h$). Similarly, we take $k_2$ to be the Fourier
mode corresponding to the component fields of the kSZ effect (ionization and
matter overdensity fields), and it probes relatively small scales ($k_2 \sim 1$
Mpc$^{-1}~h$). The goal of this description is to explain the connection between
the large-scale behavior of the 21\,cm field and the small-scale auto- or
cross-power spectra of the component fields. Though the behavior of these fields
does not perfectly map onto the behavior of the full kSZ field, they
nevertheless provide useful intuition.

During the early stages of reionization ($\ev{x_i} < 0.5$,
$8 \lesssim z \lesssim 10$ in Figure~\ref{fig:b_of_z_components}), the
inside-out nature of reionization means that ionized regions appear on small
scales near matter overdensities corresponding to areas of early galaxy
formation. Thus, regions where the 21\,cm brightness temperature is less than
the global average ($\delta T_\mathrm{21} (k_1) < \ev{T_\mathrm{21}}$) will
contain a greater-than-average number of ionized regions. The larger number of
ionized regions means that there is less neutral hydrogen, diminishing the
magnitude of the 21\,cm signal in these regions.  These same ionized regions
lead to an increase in the small-scale power of the ionization field power
spectrum $P_{xx}(k_2)$. Conversely, regions where the 21\,cm brightness
temperature is greater than the global average
($\delta T_\mathrm{21} (k_1) > \ev{T_\mathrm{21}}$) contain fewer ionized
regions, and therefore less small-scale power in the ionization field,
$P_{xx}(k_2)$. In both cases, the amplitude of the 21\,cm brightness temperature
on large scales is anticorrelated with the amplitude of the ionization power
spectrum on small scales. Therefore, we expect that the full bispectrum estimate
$B_{\mathrm{21cm},x,x}$ to have a large negative value at these redshifts due
to the anticorrelated behavior between the 21\,cm field $\delta T_\mathrm{21}$
and the ionization field power spectrum $P_{xx}$.

Using a similar line of reasoning, we can conclude that the opposite behavior
should be observed for the late stages of reionization ($\ev{x_i} > 0.5$,
$6 \lesssim z \lesssim 8$). For regions that have a local value of
$\delta T_b(k_1)$ below the global average, there are few remaining pockets of
neutral gas, which leads to a relatively low amplitude in the ionization field
power spectrum $P_{xx}(k_2)$.  Conversely, those regions with above-average
21\,cm fluctuations have more neutral regions, and therefore a larger amplitude
of the ionization field power spectrum at small scales. In other words, the
small-scale ionization power depends largely on the number of ionized regions at
early times, and the number of remaining neutral regions at late times. Taken
together, these features mean that the large-scale 21\,cm fluctuations are
positively correlated with the small-scale ionization field power spectrum
during the late stage of ionization.

Another way to look at this is as follows. The small-scale ionization field in a
region of above-average brightness temperature should resemble a ``typical''
region at an earlier time, when the \textit{global} average brightness
temperature was higher. The small-scale ionization field undergoes a ``rise''
and ``fall'' due to the changing ionization fraction of each subregion. The
sign of the correlation between brightness temperature and small-scale
ionization power then depends on whether the average ionization power spectrum
(on small scales) is an increasing or decreasing function of global average
brightness temperature. Since the ionization power spectrum undergoes a ``rise''
and ``fall'' with increasing ionization fraction (decreasing brightness
temperature), the sign of the correlation reflects whether the ionization power
spectrum is in the ``rising'' or ``falling'' stage. Near the middle of
reionization, the small-scale ionization power spectrum is a relatively flat
function of global average brightness temperature, and the correlations
discussed here are correspondingly weak.

\begin{figure}[t]
  \centering
  \includegraphics[width=0.45\textwidth]{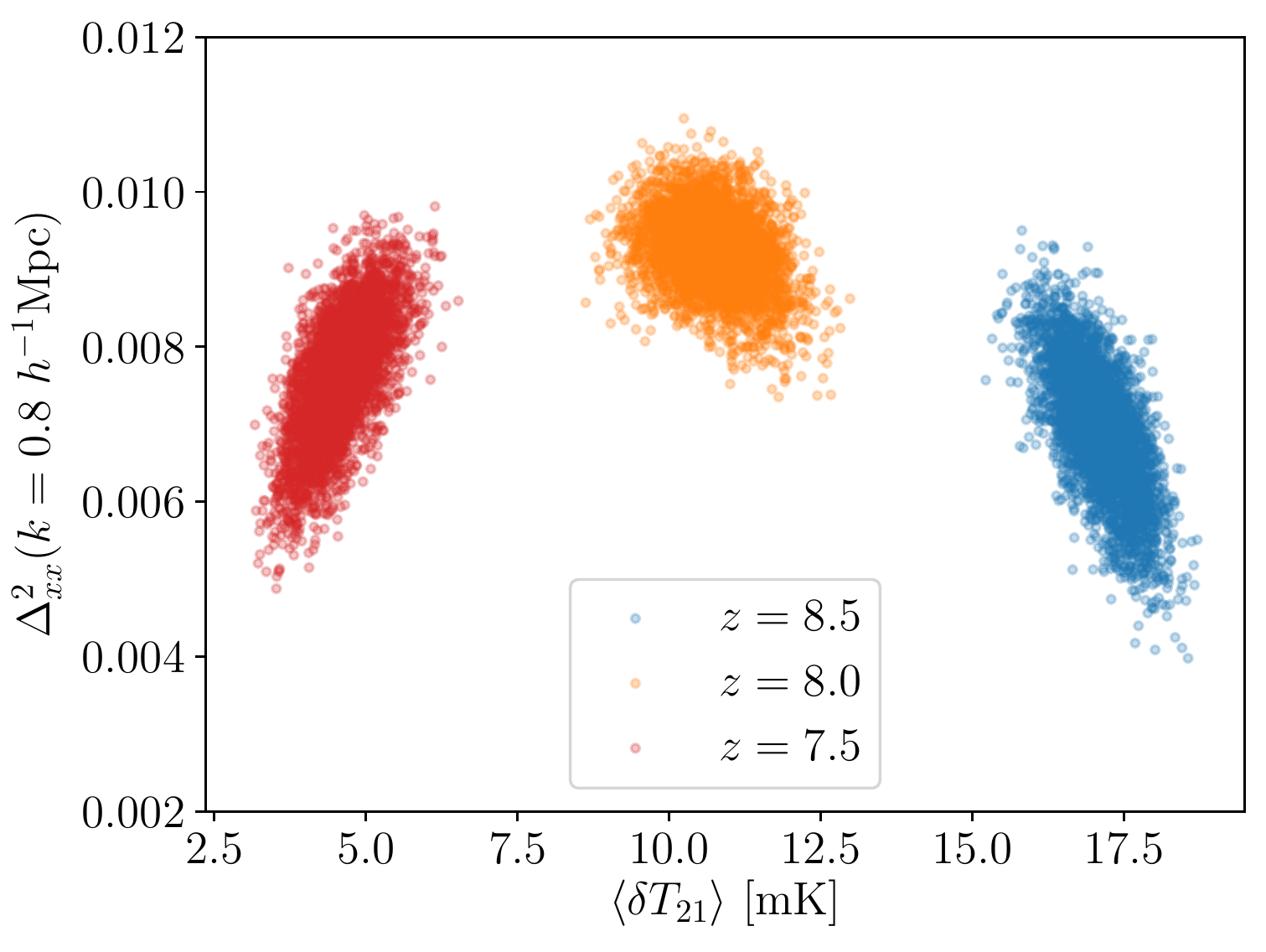}
  \caption{A scatter plot of the local average value of the 21\,cm brightness
    temperature $\delta T_b$ and the power spectrum of the ionization field
    $\Delta^2(k)$ at $k~=~0.8$~Mpc$^{-1}h$ for the fiducial reionization
    scenario. As explained in Sec.~\ref{sec:toy_model} and visualized in
    Figure~\ref{fig:tslices}, the underlying physical mechanism dictates that,
    at early times, these quantities are anticorrelated (blue points), and are
    positively correlated at late times (red points). Near the midpoint of
    reionization, there is no visible correlation (orange points).}
  \label{fig:scatter}
\end{figure}

Figure~\ref{fig:tslices} shows several 2D slices through the simulation volume
at early stages of reionization ($z = 8.5$, $\ev{x_i} \sim 0.25$) and late
stages ($z = 7.5$, $\ev{x_i} \sim 0.75$). At each of these redshifts, the 21\,cm
brightness temperature in Equation~(\ref{eqn:t21}) of sub-volumes with length
$L=125$ $h^{-1}$Mpc is visualized. We select sub-volumes where the local
average brightness temperature $\delta T_b$ is above (left column) and below
(right column) the average value for the full volume of $L=2$ $h^{-1}$Gpc. These
representative sub-volumes demonstrate the relation discussed above: at early
times, the large-scale fluctuation in $\delta T_b$ is anticorrelated with the
small-scale ionization field power spectrum, while being positively correlated
at late times.

Figure~\ref{fig:scatter} makes this behavior more quantitative. At both early
and late times, we divide the full $L=2$ $h^{-1}$Gpc volume into independent
sub-volumes of $L=125$ $h^{-1}$Mpc, leading to 4096 such sub-volumes. For each
sub-volume, we compute the average local 21\,cm brightness temperature
$\delta T_b$ and the ionization field power spectrum $P_{xx}$. To avoid
introducing artifacts associated with computing the Fourier transform on a
nonperiodic volume, we apply a cosine window to the ionization field of the
sub-volume before performing an FFT. We then compute the power spectrum at
scales of $k~=~0.8$~Mpc$^{-1}h$ for each sub-volume. We then plot the average
21\,cm brightness temperature $\delta T_b$ against the (dimensionless)
ionization power spectrum $\Delta^2_{xx}$. When examined in aggregate for the
ensemble of sub-volumes, clear trends emerge. At early times (blue points),
$\delta T_b$ and $P_{xx}$ are anticorrelated, whereas at late times (red
points), $\delta T_b$ and $P_{xx}$ are positively correlated. Near the midpoint
of reionization (orange points), there is no significant correlation between
these quantities. These trends are self-consistent with the behavior of the
$B_{\mathrm{21cm},x,x}$ line seen in Figure~\ref{fig:b_of_z_components}, and
explained in the discussion above.

The behavior of the component bispectra as a function of angle $\theta_{12}$ can
also be understood using similar lines of reasoning. Quantities whose amplitudes
are more extreme for oblique angles, such as $B_{\mathrm{21cm},\rho,\rho}$ at
$z = 7$ (see Figure~\ref{fig:b_of_theta_components}), are more sensitive to
structure along filamentary structure, similar to the matter-only bispectrum
(see Appendix~\ref{appendix:bspec}). Conversely, quantities more extreme at
$\theta_{12} \sim \pi/2$ show that the signal is sensitive to under-dense
regions. This behavior is self-consistent with the results in the above figures:
the 21\,cm-matter-matter bispectrum has a stronger response along over-dense
filaments, as with the matter-only bispectrum. The 21\,cm-ionization-ionization
bispectrum has a stronger response in under-dense regions, because at late
times, those regions preferentially have nonzero $\delta T_b$. Thus, the
behavior of the bispectrum as a function of angle is self-consistent with these
results as well.

\section{Detectability}
\label{sec:snr}

The coming decade promises the construction of instruments capable of detecting
the 21\,cm signal from the EoR at high statistical significance such as HERA and
the SKA, as well as next-generation CMB experiments for mapping the kSZ signal
such as SO and CMB-S4. Given the projected level of the signal presented in
Sec.~\ref{sec:full_bspec}, a natural question is whether we should expect to
detect this signal or not in the near future. In order to explore the
detectability of the bispectrum signal of interest, we first estimate the
expected S/N in the sample-variance limited regime. This represents the best
possible S/N that might be achieved for a given sky coverage in the case of
negligible detector noise. For simplicity, we calculate the sample variance in
the Gaussian approximation. We explore this quantity in
Sec.~\ref{sec:snr_noise_free}. Following this, we turn to the question of
instrumental noise, and whether it will be larger than the uncertainty due to
sample variance. We consider this quantity in Sec.~\ref{sec:snr_instrument}. As
a further source of observational concern, there are additional systematic
observing issues, such as the presence of foreground contamination. Such a
concern is especially acute for the 21\,cm signal, and may make actual detection
of the bispectrum challenging. We turn to potential systematic observing issues
in Sec.~\ref{sec:systematics}.

\subsection{Bispectrum Compared to Gaussian Variance}
\label{sec:snr_noise_free}

The projected bispectrum defined in Equation~(\ref{eqn:limber}) is a version of
the angular bispectrum in $\ell$-space, and so we use expressions relevant to
those quantities. We also assume that the bispectrum is observed over a
particular series of $\ell$ bins such that for bin $i$, we only consider modes
$\ell_i \leq \ell \leq \ell_{i+1}$. The Gaussian variance of the bispectrum is
then given by \citep{bucher_etal2016,coulton_spergel2019}:
\begin{widetext}
\begin{equation}
\mathrm{Var}\qty[\mathcal{B}(\ell_i,\ell_j,\ell_k)] =
\frac{1}{N_{i,j,k}^2}
\sum_{
  \substack{
    \ell_i \leq \ell_1 \leq \ell_{i+1} \\
    \ell_j \leq \ell_2 \leq \ell_{j+1} \\
    \ell_k \leq \ell_3 \leq \ell_{k+1}
  }
}
\frac{(2\ell_1 + 1)(2\ell_2 + 1)(2\ell_3 + 1)}{4\pi}
\mqty(\ell_1 & \ell_2 & \ell_3 \\ 0 & 0 & 0)^2
C_{\ell_1,\mathrm{21cm}} C_{\ell_2,\mathrm{kSZ}} C_{\ell_3,\mathrm{kSZ}} g_{\ell_1 \ell_2 \ell_3}
\label{eqn:bspec_var}
\end{equation}
\end{widetext}
where the terms in brackets are Wigner 3-$j$ symbols (which arise from
integrating products of three spherical harmonics on the celestial sphere), and
$g_{\ell_1 \ell_2 \ell_3}$ is a symmetry factor of two if $\ell_2 = \ell_3$ or
one in all other cases\footnote{In the case where the auto-bispectrum is
  computed (i.e., the three fields correspond to the same quantity), then the
  symmetry factor $g_{\ell_1 \ell_2 \ell_3}$ takes on the values of 6, 2, or 1
  for cases where three, two, or zero of the $\ell_i$ values are the same. In
  the cross-bispectrum here, the symmetry factor is only applicable when
  $\ell_2 = \ell_3$, which both correspond to the kSZ field. On the other hand,
  the sample-variance calculation for the cross-bispectrum contains additional
  terms that involve the angular cross-spectrum $C_\mathrm{21cm,kSZ}$ for cases
  where $\ell_1 = \ell_2$ or $\ell_1 = \ell_3$. Due to the wide separation in
  $\ell$-modes we use in this analysis, it is not necessary to account for these
  terms because we do not consider these combinations, which in any case are
  small at large $\ell$ due to the velocity cancellations mentioned
  above.}. $N_{i,j,k}$ is a normalization enumerating the total number of
possible triangles for a given bin, expressed as:
\begin{align}
N_{i,j,k} &\equiv
\sum_{
  \substack{
    \ell_i \leq \ell_1 \leq \ell_{i+1} \\
    \ell_j \leq \ell_2 \leq \ell_{j+1} \\
    \ell_k \leq \ell_3 \leq \ell_{k+1}
  }
}
\frac{(2\ell_1 + 1)(2\ell_2 + 1)(2\ell_3 + 1)}{4\pi} \notag \\
&\quad \times \mqty(\ell_1 & \ell_2 & \ell_3 \\ 0 & 0 & 0 )^2.
\label{eqn:ndelta}
\end{align}
The normalization $N_{i,j,k}$ measures the full number of triangles that can be
formed on the celestial sphere. To account for a finite survey area
characterized by $f_\mathrm{sky}$, we adjust the variance by dividing
Equation~(\ref{eqn:bspec_var}) by this factor. We use a value of
$f_\mathrm{sky} = 0.01$, which represents a modest overlap of experimental
footprints. Such a mutual sky-covering fraction should be feasible for HERA and
SO, which are both situated in the southern hemisphere around $-30^\circ$ of
latitude. $C_{\ell_i}$ are the 2D projected power spectra defined using the
Limber approximation for the power spectrum. For the kSZ signal, we use
Equation~(\ref{eqn:clksz}). For the 21\,cm field, we use:
\begin{equation}
C_{\ell,\mathrm{21cm}}(\ell) = \int \frac{\dd{\chi}}{\chi^2} W_\mathrm{21cm}(\chi)^2 P_\mathrm{21cm,21cm}(\ell/\chi),
\label{eqn:limber_t21}
\end{equation}
where $W_\mathrm{21cm}(\chi)$ is the 21\,cm window function defined in
Equation~(\ref{eqn:w21cm}), and $P_\mathrm{21cm,21cm}$ is the auto-power
spectrum of the 21\,cm field. To compute the variance self-consistently, one
must use the same window functions for the two-point and three-point Limber
integrals.

\begin{figure*}[t]
  \centering
  \includegraphics[width=0.32\textwidth]{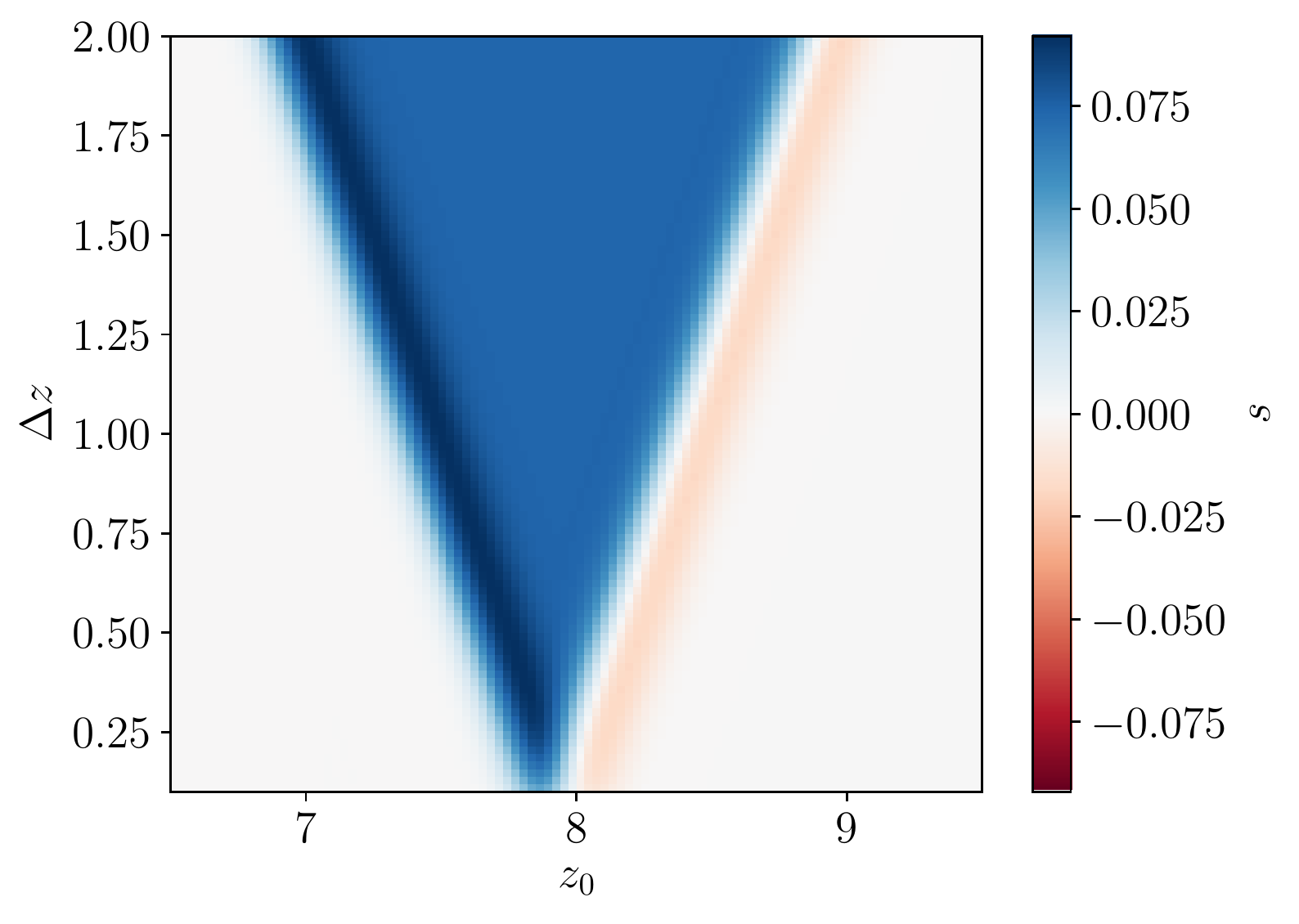} \hfill
  \includegraphics[width=0.32\textwidth]{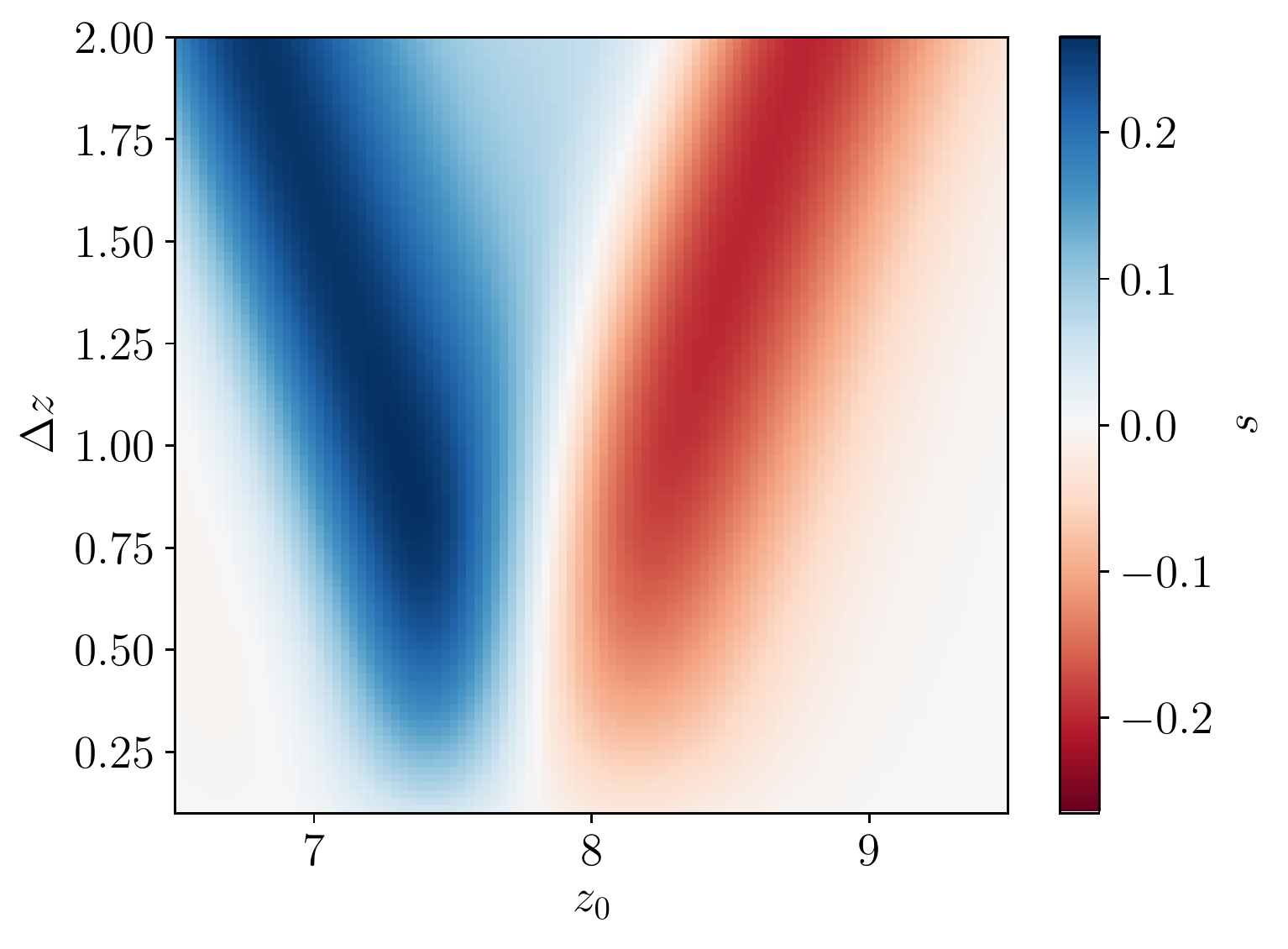} \hfill
  \includegraphics[width=0.32\textwidth]{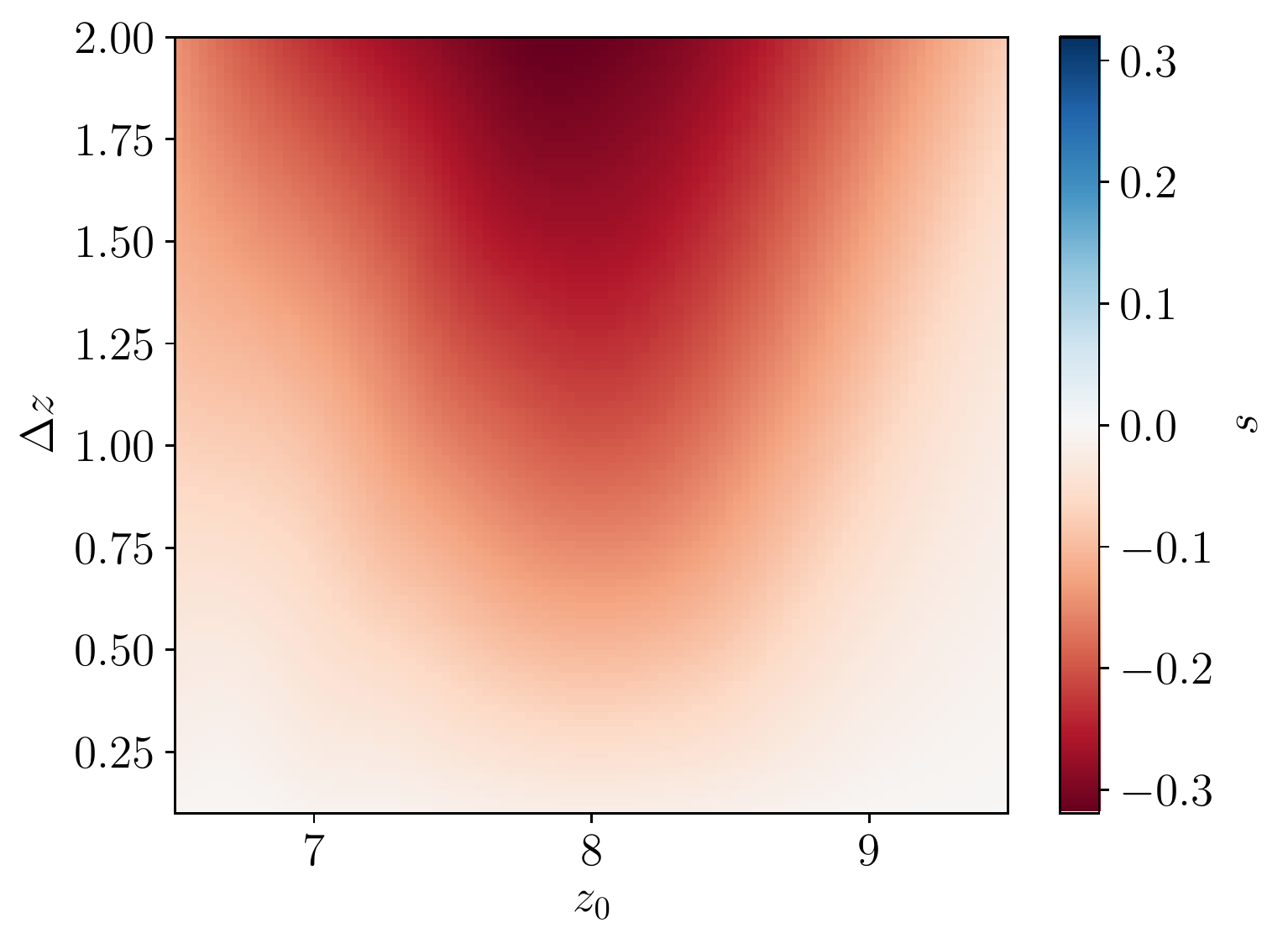}
  \caption{The ratio between the bispectrum $\mathcal{B}$ from
    Equation~(\ref{eqn:limber}) and the Gaussian variance defined in
    Equation~(\ref{eqn:bspec_var}) as a function of the top-hat 21\,cm window
    function center $z_0$ and width $\Delta z$. This quantity does not include
    experimental noise, and instead compares the signal and sample variance
    (computed in the Gaussian approximation). The combination of triangles that
    contribute to the signal in this figure represents a very small fraction of
    the total number of triangles contained within the projected survey areas
    and range of $\ell$-modes considered here. Therefore, we consider the
    cumulative S/N ratio obtainable by combining triangles across the full range
    of $\ell$-modes in Sec.~\ref{sec:cum_snr}.}
  \label{fig:snr_gaussian}
\end{figure*}

We compute the variance of the bispectrum given in
Equation~(\ref{eqn:bspec_var}) for the different 21\,cm window functions
presented in Figure~\ref{fig:bspec_limber}. For fast and efficient computation
of Wigner 3-$j$ symbols, we use the \texttt{wigxjpf} library
\citep{johansson_forssen2016}. Alternatively, in the flat-sky approximation for
$\ell_i \gg 1$, the following relationship holds \citep{joachimi_etal2009}:
\begin{equation}
\mqty(\ell_1 & \ell_2 & \ell_3 \\ 0 & 0 & 0 )^2 \approx \frac{\Lambda(\ell_1, \ell_2, \ell_3)}{2 \pi},
\label{eqn:wigner_approx}
\end{equation}
where
\begin{multline}
\Lambda(\ell_1, \ell_2, \ell_3) = \\
\frac{4}{\sqrt{2\ell_1^2\ell_2^2 + 2\ell_2^2\ell_3^2 + 2\ell_1^2\ell_3^2 - \ell_1^4 - \ell_2^4 - \ell_3^4}},
\label{eqn:lambda}
\end{multline}
subject to the same selection criteria as the original Wigner 3-$j$
symbols.\footnote{Namely, these are: (1) the parity condition that
  $\ell_1 + \ell_2 + \ell_3$ is even, and (2) the triangle inequality
  $\abs{\ell_1 - \ell_2} \leq \ell_3 \leq \ell_1 + \ell_2$. In cases where the
  triangle inequality is exactly satisfied (e.g., $\ell_1 + \ell_2 = \ell_3$),
  the approximation yields division by 0 in Equation~(\ref{eqn:lambda}), whereas
  the Wigner 3-$j$ symbol is nonzero. We use the exact expression for these
  combinations.}  We have verified that the above approximation yields values
that are better than 1\% accurate for typical $\ell$ combinations considered in
this work. We use the approximate form in Equation~(\ref{eqn:wigner_approx})
when all $\ell_i \geq 200$, and the exact form from \texttt{wigxjpf} otherwise.

\subsubsection{Single-window Signal-to-noise}
\label{sec:single_snr}

Equation~(\ref{eqn:bspec_var}) is defined for a particular combination of
$\ell$-mode ranges for each of the $\ell_i$ values. To make a prediction of the
individual sensitivity for a particular $\ell$-range, we define ranges for each
of the $\ell_i$ values. As an example for illustrative purposes, we choose
$75 \leq \ell_1 \leq 105$ for the 21\,cm modes of interest and
$2950 \leq \ell_{2,3} \leq 3050$ for the kSZ modes. We also compute the $C_\ell$
spectrum for each 21\,cm window function self-consistently, as in the
computation of $\mathcal{B}$ in Figure~\ref{fig:bspec_limber}. After computing
the variance $\mathrm{Var}(\mathcal{B})$, we take the square root to obtain the
standard deviation, and compare that value with the bispectrum $\mathcal{B}$ as
computed by Equation~(\ref{eqn:limber}). We express this dimensionless quantity
$s$ as:
\begin{equation}
s(\ell_1,\ell_2,\ell_3) \equiv \frac{\mathcal{B}(\ell_1,\ell_2,\ell_3)}{\sqrt{\mathrm{Var}[\mathcal{B}(\ell_1,\ell_2,\ell_3)]}}.
\label{eqn:snr}
\end{equation}

Figure~\ref{fig:snr_gaussian} shows this quantity $s$ for the short, fiducial,
and long reionization histories. As in Figure~\ref{fig:bspec_limber}, we have
computed this quantity for different top-hat 21\,cm window functions
parameterized by their center $z_0$ and width $\Delta z$. Interestingly,
although the intrinsic signal $\mathcal{B}$ is largest for the narrowest 21\,cm
windows, the quantity $s$ increases for larger windows. Even though wider 21\,cm
windows lead to a smaller signal, they also decrease the amplitude of
$C_{\ell,\mathrm{21cm}}$ by a larger amount. Essentially, the top-hat window
acts as an average of the 21\,cm field combined from different redshifts. This
decoheres the signal compared to a narrower window, yet preserves the
cross-correlation with the kSZ signal. The end result is that for the 21\,cm
window functions considered here, a larger window function leads to a higher S/N
detection, although the signal has a less straightforward interpretation in this
case. Specifically, the values of $\{(z_0, \Delta z)\}$ which maximize the
response for the short, fiducial, and long histories are
$\{(7.8, 0.39), (7.3, 0.94), (7.8, 2.0)\}$, respectively. The general trend of
larger values of $\Delta z$ leading to higher S/N values as the duration of
reionization becomes larger is related to the fact that the kSZ visibility
function $g(\chi)$ has broader support in redshift space as the duration
increases.

Note that the magnitude of the quantity $s$ is less than one for all choices of
the 21\,cm window function. This result means that the bispectrum signal is
smaller than the Gaussian variance for an individual $\ell$-range, though it is
possible to combine the significance across different $\ell$-ranges (and
multiple redshift windows) to increase the overall detectability of the
signal. We expand on this discussion below in Sec.~\ref{sec:cum_snr}.

\subsubsection{Cumulative Signal-to-noise}
\label{sec:cum_snr}

The above discussion in Sec.~\ref{sec:single_snr} examined the ratio between the
measured bispectrum and Gaussian variance for a single combination of
$\ell$-ranges relevant to upcoming observations. However, in principle one can
combine the significance from multiple $\ell$-ranges to yield a highly
significant detection above the level of Gaussian variance. To understand the
total sensitivity from combining measurements from HERA and SO, we expand the
range of $\ell$-modes considered: we examine all modes
$90 \leq \ell_\mathrm{21cm} \leq 1000$ and
$3000 \leq \ell_\mathrm{kSZ} \leq 6000$. These scales represent modes accessible
by HERA and SO, respectively, and should be measured at high significance
relative to instrumental noise (see further discussion below in
Sec.~\ref{sec:snr_instrument}). We use bin widths of
$\Delta \ell_\mathrm{21cm} = 30$ and $\Delta \ell_\mathrm{kSZ} = 100$, so that
the relative width in $\ell$-space is comparable between the two signals. Due to
the triangle inequality enforced in Equations~(\ref{eqn:bspec_var}) and
(\ref{eqn:ndelta}), there will be some combinations of $\ell$-ranges that do not
yield valid triangles (e.g., $\ell_\mathrm{21cm} = 90$,
$\ell_\mathrm{kSZ,1} = 3000$, $\ell_\mathrm{kSZ,2} = 6000$). Nevertheless, there
are many valid combinations that can be formed, and so the cumulative
significance can be increased substantially.

For each combination of $\ell$-bins that contains valid triangles, we compute
the variance $\mathrm{Var}(\mathcal{B})$ according to
Equation~(\ref{eqn:bspec_var}). Computing the bispectrum for each window is
computationally prohibitive, and so we have computed the bispectrum explicitly
for the combinations of:
$\{(\ell_\mathrm{21cm} = 90, \ell_\mathrm{kSZ} = 3000), (\ell_\mathrm{21cm} =
90, \ell_\mathrm{kSZ} = 6000), (\ell_\mathrm{21cm} = 1000, \ell_\mathrm{kSZ} =
3000),$ and $(\ell_\mathrm{21cm} = 1000, \ell_\mathrm{ksz} = 6000)\}$. We
compute the value of $\mathcal{B}$ according to Equation~(\ref{eqn:limber}), and
linearly interpolate the resulting function in log-space to yield an approximate
value for $\mathcal{B}$ at the particular combination of central $\ell_i$
values. We have computed the value of $\mathcal{B}$ at several points interior
to the convex hull defined by these four combinations, and verified that this
interpolation produces values that are typically accurate to $\sim 50$\%. We
then combine the significance from different windows in quadrature to get
$s_\mathrm{cum}$:
\begin{equation}
s_\mathrm{cum}^2 = \sum_{\ell_{i,j,k}} \qty(\frac{\mathcal{B}(\ell_i, \ell_j, \ell_k)}{\sqrt{\mathrm{Var}\qty[\mathcal{B}(\ell_i,\ell_j,\ell_k)]}})^2.
\label{eqn:s_cum}
\end{equation}
To demonstrate how this cumulative significance depends on the maximum value of
$\ell_\mathrm{21cm,max}$, we compute this quantity for all combinations of
$3000 \leq \ell_\mathrm{kSZ} \leq 6000$ and
$90 \leq \ell_\mathrm{21cm} \leq \ell_\mathrm{21cm,max}$. Such a calculation
makes clear which observational modes from HERA are most important for
increasing the overall sensitivity of the statistic.

\begin{figure}
  \centering
  \includegraphics[width=0.45\textwidth]{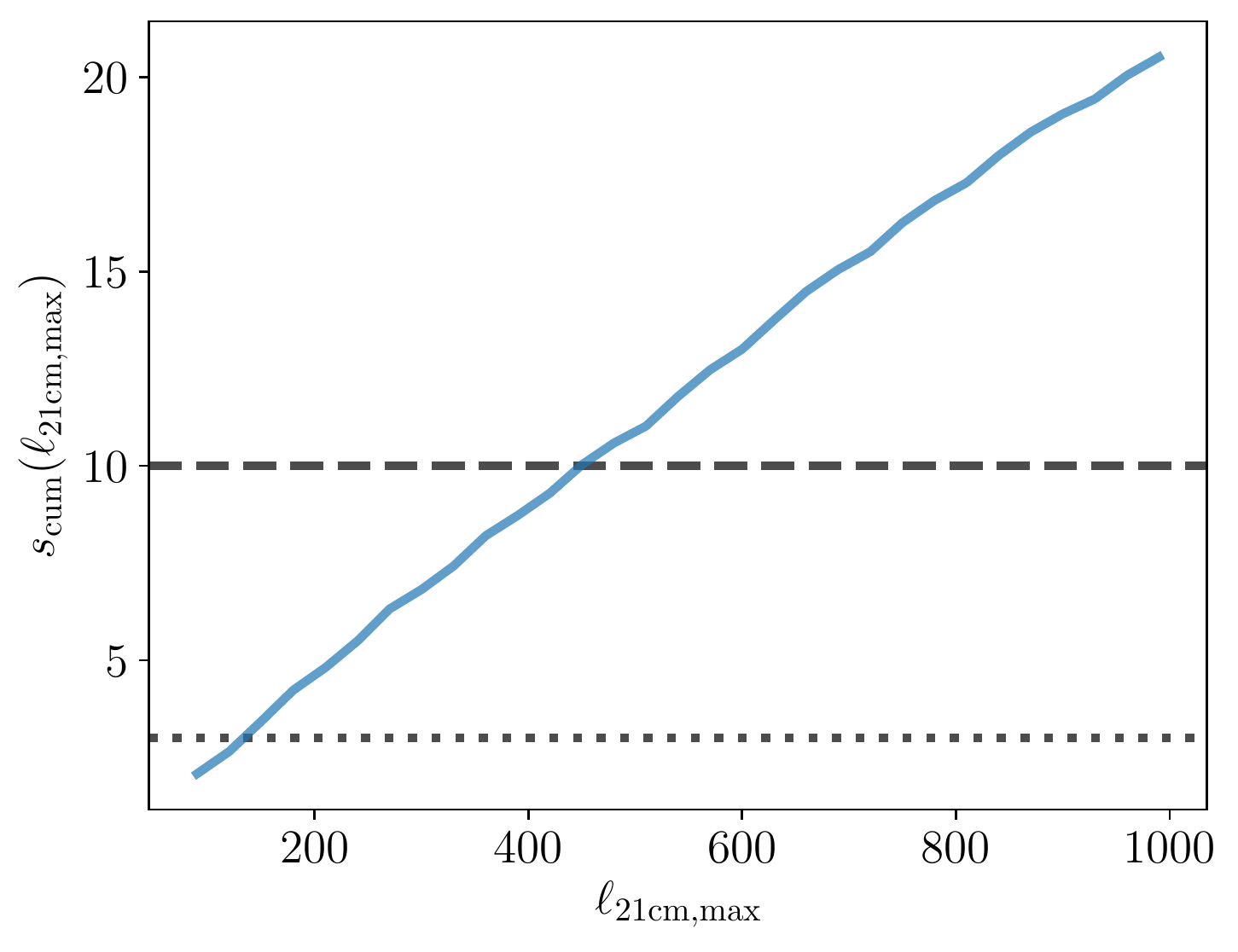}
  \caption{The cumulative S/N ratio of the bispectrum compared to the sample
    variance defined in Equation~(\ref{eqn:s_cum}) as a function of the maximum
    21\,cm $\ell$-mode used, when summing over all valid combinations of kSZ
    modes $3000 \leq \ell_\mathrm{kSZ} \leq 6000$ for a single redshift window
    defined by $W_\mathrm{21cm}(\chi)$ with parameters $z_0 = 7.5$ and
    $\Delta z = 1$. The dotted line shows a 3$\sigma$ detection, and the dashed
    line shows a 10$\sigma$ detection. As discussed in
    Sec.~\ref{sec:snr_instrument}, the instrumental noise is projected to be
    below these values. However, systematic effects, particularly foreground
    contamination of the 21\,cm signal, may lead to lower significance in
    practice. See Sec.~\ref{sec:systematics} for further discussion.}
  \label{fig:cum_snr}
\end{figure}

Figure~\ref{fig:cum_snr} shows the cumulative significance defined in
Equation~(\ref{eqn:s_cum}) as a function of $\ell_\mathrm{21cm,max}$. A dotted
line shows where $s_\mathrm{cum} = 3$, meaning the bispectrum can be measured at
a 3$\sigma$ significance. This threshold is crossed for
$\ell_\mathrm{21cm,max} \sim 180$, meaning that relatively few $\ell$-modes from
HERA would be necessary to make a significant detection. If all modes up to
$\ell_\mathrm{21cm,max} = 1000$ are included, then a roughly 20$\sigma$
detection is statistically possible. The results of \citet{ma_etal2018} are
broadly consistent with the cumulative S/N found here, suggesting that both
quantities have similarly large statistical significance. Note that this
calculation does not consider the potential impact of systematic errors related
to observations. We discuss potential issues further below in
Sec.~\ref{sec:systematics}.

The forecast above of a 20$\sigma$ detection relies only on a single tomographic
redshift bin centered on $z_0 = 7.5$ with a width of $\Delta z = 1$. Given the
S/N shown in Figure~\ref{fig:snr_gaussian}, we expect detections of several
$\sigma$ might be possible in the absence of foreground contamination using an
additional redshift bin centered on $z_0 = 8.5$, or even $z_0 = 9.5$ with a
width of $\Delta z = 2$. It may also be feasible to use narrower windows with
$\Delta z = 0.5$, and construct several different nonoverlapping observation
windows that yield statistically significant detections. In principle, the
21\,cm--kSZ--kSZ bispectrum might therefore allow some ability to tomographically
reconstruct the global reionization history.

\vspace{20pt}

\subsection{Experimental Noise}
\label{sec:snr_instrument}

As mentioned above, a real-world measurement of the bispectrum will have to
contend with both the sample variance of the signal as well as uncertainty
introduced by the detectors. When accounting for both sources simultaneously,
the mathematical form of Equation~(\ref{eqn:bspec_var}) must include both
sources. Specifically, the terms including the angular power spectra
$C_{\ell_i}$ should include both the signal (for the sample variance) and the
detector noise, which we denote $N_{\ell_i}$. For example, the first term should
be substituted
$C_{\ell_1,\mathrm{21cm}} \to (C_{\ell_i,\mathrm{21cm}} +
N_{\ell_1,\mathrm{21cm}})$, with analogous changes for the other two terms. In
the limit that $N_{\ell_i} \ll C_{\ell_i}$, then Equation~(\ref{eqn:bspec_var})
reduces to the form given above. However, this assumes that the noise is
subdominant term-by-term for each $\ell$ mode considered. We will examine the
projected noise behavior of 21\,cm detectors such as HERA, followed by CMB
detectors such as SO.

\subsubsection{21\,cm Noise Spectra}
\label{sec:snr_21cm}

Many of the current- and next-generation 21\,cm experiments are radio
interferometers. Individual baselines are sensitive to a specific $\ell$-mode in
the sky at a given frequency, and measurements from multiple baselines can be
combined to generate an image. Rather than estimating the noise on the image as
a whole, we derive here an expression for the instrumental noise when measuring
a single baseline, which is the relevant quantity for a measurement of the
bispectrum. Thus, as a noise model for the 21\,cm signal for a single baseline
$N_{\ell,1}$, we use the following expression from
\citet{zaldarriaga_etal2004}:
\begin{equation}
N_{\ell,1} = \frac{T_\mathrm{sys}^2(2\pi)^2}{\Delta \nu t_\nu \dd[2]{\ell}},
\label{eqn:cnoise_t21}
\end{equation}
where $T_\mathrm{sys}$ is the system temperature, $\Delta \nu$ is the bandwidth
of the observation, $t_\nu$ is the total amount of coherent time observing a
single Fourier pixel over a season, and $\dd[2]{\ell} \equiv (\Delta \ell)^2$
denotes the $\ell$-range being observed.

\begin{figure}
  \centering
  \includegraphics[width=0.45\textwidth]{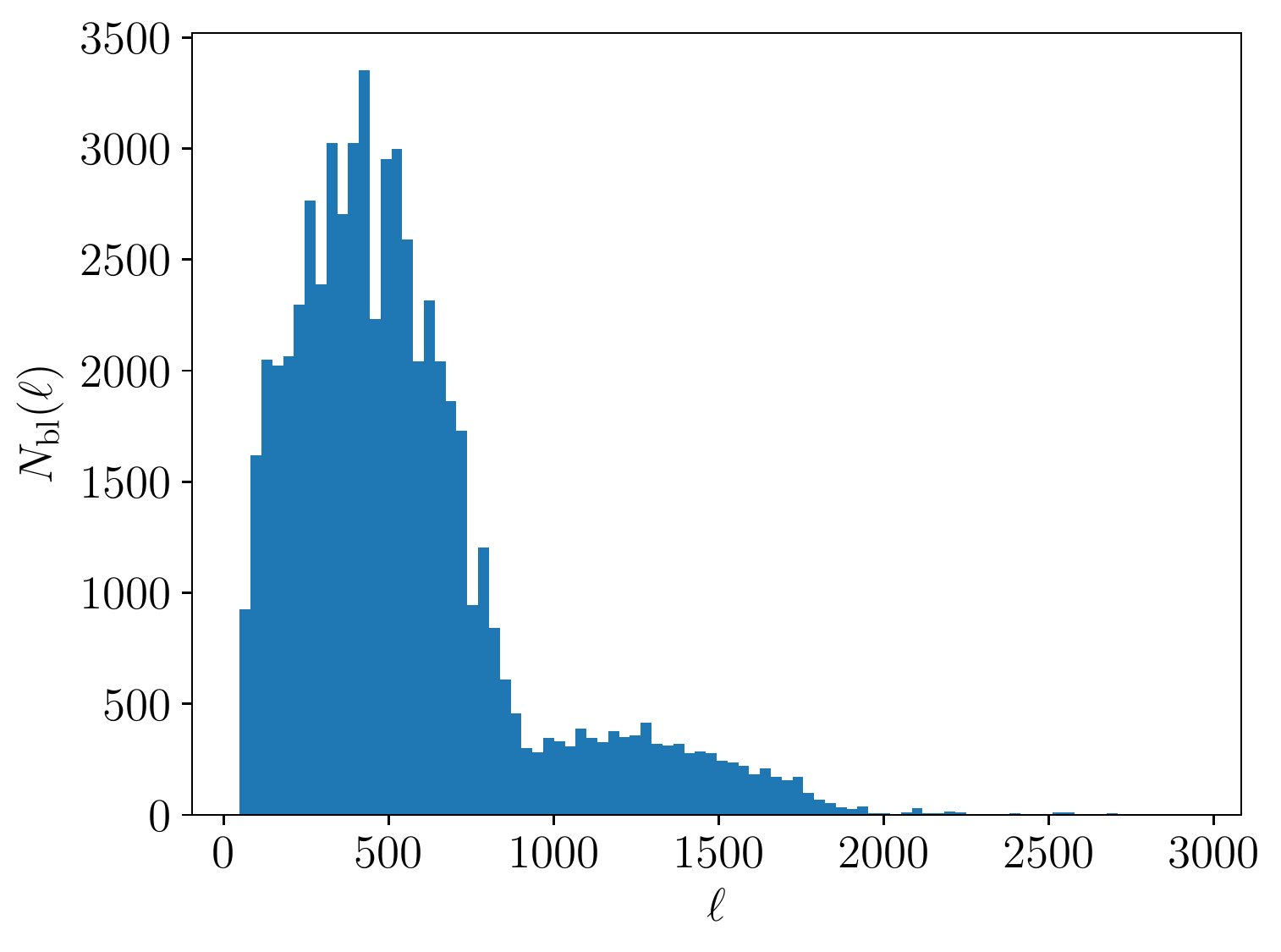}
  \caption{The number of baselines in the full 350-element HERA array that
    observe a given $\ell$-mode at $z = 8$. This quantity appears in
    Equation~(\ref{eqn:cnoise_t21_total}) and assumes a flat-sky observation
    where baselines are stationary in the $uv$-plane.}
  \label{fig:nbl_dist}
\end{figure}

When forecasting this quantity for HERA, we assume that each baseline will be
observed and added coherently for a time $t_c$ across each night for a total of
$N_\mathrm{obs}$ nights in a single observing season. Using this observing
strategy, we use $t_\nu = N_\mathrm{obs}t_c$ in Equation~(\ref{eqn:cnoise_t21}),
meaning that the noise for a single baseline averages down linearly in time over
the coherence time scale. We also assume that we incoherently average the signal
from $N_i$ coherently averaged time windows. Furthermore, HERA features many
nominally redundant baselines, which are probing the same or statistically
equivalent modes on the sky. The noise for the entire array will be reduced
linearly by the total number of baseline pairs for a given $\ell$-mode
$N_\mathrm{bl}(\ell)$, as well as the square root of the number of incoherently
averaged time windows $N_i$. Thus, the noise level for the full array is:
\begin{equation}
  N_{\ell,\mathrm{tot}} = \frac{N_\mathrm{\ell,1}}{N_\mathrm{bl}(\ell) \sqrt{N_i}}.
\label{eqn:cnoise_t21_total}
\end{equation}
To compute the quantity $N_i$, we use the total length of observing $t_i$
divided by the coherently averaging time length $t_c$: $N_i = t_i / t_c$. We
assume that the coherent integration time is $t_c = 15$ minutes, and the total
observing window $t_i$ is 8 hr, giving $N_i = 32$. We also assume an observing
season of 100 days, so $N_\mathrm{obs} = 100$. As above in
Sec.~\ref{sec:snr_noise_free}, we assume that $\Delta \ell = 30$, which comes
from the product of the HERA primary beam and the fringe term for a 14.6 m
baseline (the shortest baseline in the HERA array) and a wavelength of
$\lambda = 2$ meters. We use $T_\mathrm{sys} = 400$ K, a value consistent with
estimates across different frequency ranges and LST windows (HERA Public Memo
\#19\footnote{\url{http://reionization.org/wp-content/uploads/2017/04/HERA19_Tsys_3April2017.pdf}}).
We use an observing bandwidth of $\Delta \nu = 17$ MHz, corresponding to a
window of width $\Delta z = 1$ centered at $z = 8$. Given the highly redundant
design of HERA, a significant number of baselines can be constructed that are
sensitive to modes $90 \lesssim \ell \lesssim 1000$. Generally, a given $\ell$
mode in this range is simultaneously observed by roughly 1,000 baselines. (See
Figure~\ref{fig:nbl_dist} for the full distribution.) Using these quantities in
Equation~(\ref{eqn:cnoise_t21_total}), we find:
\begin{align}
N_{\ell,\mathrm{tot}} &= \frac{T_\mathrm{sys}^2 (2\pi)^2}{\Delta \nu N_\mathrm{obs} t_c(\Delta \ell)^2 N_\mathrm{bl}(\ell) \sqrt{N_i}} \notag \\[0.5em]
&= 0.81 \qty(\frac{1000}{N_\mathrm{bl}(\ell)}) (\mu\mathrm{K} \cdot \mathrm{rad})^2.
\end{align}
This quantity represents the nominal noise sensitivity of HERA with 1000
baselines measuring a given $\ell$-mode for a whole observing season. The
amplitude of the corresponding $C_\ell$ mode computed from
Equation~(\ref{eqn:limber_t21}) is larger by about two orders of magnitude for
the same $\ell$-range. For modes where $\ell \gtrsim 1000$, the fewer number of
redundant baselines suggests that $N_\ell \sim C_\ell$. However, as shown in
Figure~\ref{fig:cum_snr}, a significant detection can be made using only the
modes that are well-sampled by HERA.

With measurements from SKA, the projected S/N per mode is much larger than that
of HERA \citep{koopmans_etal2015}. Given this feature, the requirement that
$N_\ell \ll C_\ell$ will easily be satisfied. Once this requirement is
satisfied, the only remaining question is the size of the planned survey. A
medium-depth survey is planned for SKA Phase 1, which would cover 1000 deg$^2$
\citep{koopmans_etal2015}, which is comparable to the total survey area of
HERA. As discussed above, we assume $f_\mathrm{sky} = 0.01$, which means that
only about half of this survey would be required to overlap with the target CMB
survey. A shallow survey is also planned to cover 10,000 deg$^2$
($f_\mathrm{sky} \sim 0.25$), which will almost certainly overlap with several
upcoming CMB observations.

\subsubsection{kSZ Noise Spectra}
\label{sec:snr_ksz}

Now we turn to the noise $N_{\ell,\mathrm{kSZ}}$ compared to the signal
$C_{\ell,\mathrm{kSZ}}$. When generating a $C_\ell$ spectrum from a map of the
kSZ field, such as the one shown in Figure~\ref{fig:maps}, we find that
$D_\ell = \ell^2 C_\ell/(2\pi)$ has an amplitude of $\sim$3 $\mu$K$^2$, and is
relatively flat over $3000 \lesssim \ell \lesssim 6000$ in
$D_\ell$-space\footnote{Note that this represents the patchy contribution to the
  total kSZ signal. The low-redshift, post-reionization contribution is
  comparable in magnitude, and so the sample-variance contribution may need to
  be adjusted by up to a factor of two in the above analysis.}. This is
consistent with previous theoretical investigations of the patchy kSZ effect
\citep{battaglia_etal2013b,alvarez2016}. This result is also consistent with the
recent measurement of the SPTPol + SPT-SZ surveys \citep{reichardt_etal2020}.

We begin by using
``BoloCalc'',\footnote{\url{https://github.com/chill90/BoloCalc/}} a software
tool for estimating the sensitivity of SO \citep{hill_etal2018}. We assume an
observing time of one year, with an observational efficiency of 20\% (i.e.,
about 2.5 months total observing time). Using a sky coverage of
$f_\mathrm{sky} = 0.01$, simulation with the Large-Aperture Telescope (LAT) for
the baseline observation assumptions yields a map depth $\sigma_S$ of 2.29
$\mu$K arcmin. The noise induced by the LAT beam is relatively flat in
$\ell$-space between $3000 \leq \ell \leq 6000$ \citep{so_science_goals}, so the
effect of the beam is minor over this range. To convert from map depth to noise
spectra in $\ell$-space, we use the relation \citep{knox1995,staggs_etal2018}:
\begin{align}
N_\ell &= \sigma_s^2 \exp(\frac{\ell (\ell + 1) \theta_\mathrm{FWHM}^2}{8 \ln 2}) \qty(\frac{\pi}{10800})^2 \label{eqn:cnoise_ksz}\\[0.5em]
\Rightarrow N_{3000} &= 5.09 \times 10^{-7} (\mu\mathrm{K}\cdot \mathrm{rad})^2,\notag
\end{align}
where $\theta_\mathrm{FWHM}$ is the FWHM for the LAT beam in radians. This value
changes as a function of observing frequency, but is about 1 arcmin in size for
the bands relevant to the kSZ measurement. We use a value of
$\theta_\mathrm{FWHM} = 1\arcmin$ for the purposes of the above
calculation. Converting to $D_\ell$, this is an equivalent noise of
$D_{3000} = 0.73$ $\mu$K$^2$. This quantity is below the expected level of the
signal, and so satisfies the requirement that
$N_{\ell,\mathrm{kSZ}} < C_{\ell,\mathrm{kSZ}}$. However, at $\ell = 6000$,
$D_{6000} = 4.4$ $\mu$K$^2$, and so the detector noise is comparable to or
potentially larger than the signal. This makes the above sample-variance limit
optimistic. The noise can be made smaller through additional observation, though
the sensitivity only increases as the square root of the amount of time
observed. In practice, systematic observing issues may be more difficult to
overcome.

Although making a measurement with the required level of sensitivity may require
several months of observation, the entire footprint of HERA is already planned
to be measured as part of the main $f_\mathrm{sky} = 0.4$ survey. The smaller
patch with $f_\mathrm{sky} = 0.01$ can be made in addition to the main
measurement, meaning that less than 2.5 months of additional observation would
be required. In addition to detecting the bispectrum measurement proposed above,
this deep field could be used to measure the kSZ auto-power spectrum to high
certainty, providing an additional observational measurement related to the
EoR. The results from the power spectrum measurement can be used in conjunction
with the bispectrum measurement to provide additional cross-checks of results
from HERA alone.

Future CMB experiments beyond SO are projected to have even better map
sensitivities. One in particular is the Probe of Inflation and Cosmic Origins
(PICO), a proposed space-based mission for providing an all-sky map of the CMB
to very high sensitivity. The target map sensitivity is $\sigma_S = 0.87$
$\mu$K arcmin \citep{pico2019}, which is more than factor of two better than the
SO sensitivity (though with a slightly larger beam size). As a result, the
projected noise sensitivity as calculated by Equation~(\ref{eqn:cnoise_ksz})
will be better than that of SO by a modest amount. At the same time, PICO will
provide a map of the full sky, meaning there is significant opportunity for
cross-correlation with large-area future 21\,cm surveys such as SKA.

\subsection{Systematic Observing Issues}
\label{sec:systematics}

The discussion above in Sec.~\ref{sec:snr_instrument} only captures the
statistical uncertainty associated with measuring the power spectrum and
bispectrum. There are systematic sources of error and uncertainty associated
with both the 21\,cm and kSZ observations, which may make cross-correlation
measurements in the future difficult. One obvious source of systematic error not
quantified in the discussion above is the foreground ``wedge'' of 21\,cm
observations
\citep{datta_etal2010,morales_etal2012,vedantham_etal2012,liu_etal2014a}. When
working with 21\,cm observations, the observational modes can be written as
$k_\parallel$ and $k_\perp$, where $k_\parallel$ measures the power along the
line of sight, and $k_\perp$ denotes a Fourier mode in the plane of the
sky. These $k_\parallel$ modes measure the response along the frequency
direction. Bright foreground emission (largely synchrotron emission from the
Milky Way) is smooth as a function of frequency, and so appears as significant
power at small $k_\parallel$ observational modes. The chromaticity of the
interferometer scatters power from these small $k_\parallel$ modes to larger
ones, with the contamination reaching larger $k_\parallel$ modes for larger
values of $k_\perp$. In particular, the $k_\parallel = 0$ mode is always
contaminated by bright foregrounds, and will almost certainly have to be removed
when analyzing the 21\,cm signal. The kSZ signal is an integrated quantity
appearing in the 2D CMB, and is only sensitive to $k_\parallel = 0$ modes. Due
to the fact that the bispectrum must be measured from closed triangles, and each
component must individually sum to zero, measuring the bispectrum in practice
implies that the $k_\parallel = 0$ mode for the 21\,cm field must be used.

To measure the statistic discussed here, a different approach to foreground
cleaning will need to be implemented, rather than the usual foreground wedge
avoidance. Instead, it will be necessary to pursue efforts along the lines
proposed by, e.g., \citet{zaldarriaga_etal2004} for measuring the 21\,cm angular
power spectrum. The basic idea here is that if the foregrounds are highly
correlated across frequency, one can use this property to separate the signal
and foregrounds while retaining some purely transverse modes. For instance,
redshifted 21\,cm measurements at post-reionization frequencies may provide
foreground templates.

We defer a full investigation to future work, but we can give some quantitative
indication of how spectrally smooth the foregrounds need to be for this approach
to be effective (see, e.g., \citealt{zaldarriaga_etal2004}). For simplicity,
consider two frequencies: one, $\nu'$, corresponding to a post-reionization
frequency and another, $\nu$, at the redshifted 21\,cm frequency of
interest. The key quantity here is the correlation coefficient, $r$, between the
foreground fluctuations at the two frequencies. This is defined by
\begin{equation}
r = \frac{C_{\ell,\mathrm{foreground}}(\nu,\nu')}{\sqrt{C_{\ell,\mathrm{foreground}}(\nu,\nu) C_{\ell,\mathrm{foreground}}(\nu',\nu')}},
\end{equation}
where $C_{\ell,\mathrm{foreground}}(\nu,\nu')$ gives the cross-spectrum between
the foregrounds in the two maps, while $C_{\ell,\mathrm{foreground}}(\nu,\nu)$
and $C_{\ell,\mathrm{foreground}}(\nu',\nu')$ give the respective foreground
auto-spectra. Using the high-frequency map as a foreground template, the
residual foreground power at the target frequency $\nu$ is
$C_{\ell,\mathrm{residual}} = (1 - r)C_{\ell,\mathrm{foreground}}$.

In this case, using present estimates of $C_{\ell,\mathrm{foreground}}$ from an
all-sky model of the radio sky \citep{zheng_etal2017}, the correlation
coefficient must satisfy $r \geq (1 - 10^{-8})$ for the foreground residual
contribution to the variance to be less than the signal power (e.g.,
\citealt{zaldarriaga_etal2004}). While this estimate is instructive, note that
it considers only two frequency bands. In practice, one can consider a full set
of additional spectral channels that would yield additional discriminating
power. Future work will be required to determine the correlation coefficients
between the foregrounds at different frequencies and to quantify the full
multichannel prospects here.

When considering the kSZ signal in the above analysis, we focused exclusively on
the contribution to the signal at high redshift relevant to the EoR
($z \gtrsim 6$). There is also a low-redshift contribution to the kSZ signal
from nearby galaxies. At $\ell \sim 3000$ the signal is expected to be dominated
by the EoR contribution, though the low-redshift contribution will still be
present. Optimistically, the low-redshift component will not cross-correlate
with the structures probed by the 21\,cm signal, and will merely add to the
uncertainty of the signal through incoherent noise. However, isolating the kSZ
signal to the requisite level may require more precise component separation than
has been done previously. Another promising approach for extracting the kSZ
signal is to consider higher-point functions of the CMB maps alone
\citep{smith_ferraro2017}. As with the 21\,cm signal, careful analysis and novel
techniques may be required to ensure a successful detection.

\section{Conclusion}
\label{sec:conclusion}

In this work, we explore the bispectrum between the 21\,cm and kSZ fields during
reionization in the squeezed-triangle limit. We show that the signal is very
sensitive to the duration of the universe's reionization history, depending on
both the timing of reionization as well as its duration. For our short,
fiducial, and long histories, the correlation between features appears near the
midpoint of reionization. Thus, it may be possible to confirm certain features
such as the midpoint of reionization from this statistic if known from other
observations (e.g., the 21\,cm auto-power spectrum), or to infer it by comparing
measurements with a suite of different reionization models.

As discussed in Sec.~\ref{sec:snr_noise_free}, we project that the cumulative
sample-variance limited bispectrum is detectable at more than 20$\sigma$ for a
joint measurement between HERA and SO given a sky-covering fraction of
$f_\mathrm{sky} = 0.01$ and projected noise parameters. At the same time, there
are sources of systematic uncertainties, such as how best to handle the bright
foreground contamination of 21\,cm observations. We suggest one possible
approach for mitigating these issues in Sec.~\ref{sec:systematics}, though
additional work is required to demonstrate that such an approach can remove the
foregrounds without destroying the statistical information necessary for
detecting the bispectrum.

As an alternative to the statistic presented here, future work may examine a
four-point statistic based on the 21\,cm--21\,cm--kSZ--kSZ trispectrum. Such a
statistic may not suffer from the foreground wedge contamination for
low-$k_\parallel$ modes, as 21\,cm $k$-modes can be chosen with equal and
opposite values of $k_\parallel$ far from zero. However, the trispectrum
presents additional computational challenges and detailed forecasts are required
to quantify its detectability. That said, it may be possible to extract some
four-point cross-correlation information without using the full trispectrum. One
possibility is to compute the cross-power between the squared fields
$\delta T_\mathrm{kSZ}^2$ and $\delta T_\mathrm{21cm}^2$ (after filtering out
contaminating modes, such as the primary CMB for the kSZ field). Such a
statistic may be sensitive to some of the joint information contained in the
fields, without necessitating fully computing the trispectrum. In future
studies, cross-correlating the 21\,cm and kSZ signals from the EoR may provide a
valuable cross-check on quantities inferred from either signal alone, and
represents a rich opportunity for joint-analysis in the future.

\acknowledgments{We thank Marcelo Alvarez and Charlie Hill for useful
  discussions about this project. This material is based upon work supported by
  the National Science Foundation under grant No. 1636646, the Gordon and Betty
  Moore Foundation, and institutional support from the HERA collaboration
  partners. HERA is hosted by the South African Radio Astronomy Observatory,
  which is a facility of the National Research Foundation, an agency of the
  Department of Science and Technology. This work was supported by the Extreme
  Science and Engineering Discovery Environment (XSEDE), which is supported by
  National Science Foundation grant No. ACI-1548562
  \citep{xsede2014}. Specifically, this work made use of the Bridges system,
  which is supported by NSF award No. ACI-1445606, at the Pittsburgh
  Supercomputing Center \citep{bridges2015}.}


\appendix

\section{Validation of Bispectrum Estimator}
\label{appendix:bspec}

The results in Sec.~\ref{sec:results} rely heavily on computing the bispectrum
of three fields. Computing the bispectrum is nontrivial, and it can be
computationally intensive to exhaustively compute all possible combinations of
triangle lengths $k_1,k_2,k_3$. The na\"\i ve approach formally requires a 9D
nested do-loop to construct all triangles available in a given simulation
volume, though this requirement can be reduced to a 6D loop given the closure
requirement. Nevertheless, exhaustively computing all such combinations is
computationally infeasible.

As an alternative, we implement a bispectrum estimator based on fast Fourier
transforms (FFTs). This method of computing higher-point estimators was
introduced by \citet{jeong2010}, and applied to computing the auto-bispectrum of
the 21\,cm field in \citet{watkinson_etal2017} and
\citet{majumdar_etal2018}. Briefly, this approach involves extracting particular
$\vb{k}$-modes for a given triangle (or group of triangles) characterized by the
set of $k_i$ and building associated unit-weight fields to use for
normalization. For a $p$-point estimator, this approach requires $2p$ additional
fields stored in memory, as well as applying FFTs to them. However, due to the
overall computational complexity being dominated by FFT operations, the time
requirement scales as $\mathcal{O}(N\log N)$ for a volume of $N$ elements rather
than the $\mathcal{O}(N^2)$ scaling of the na\"\i ve approach. The computational
savings become even more significant for higher-$p$ estimators, such as the
trispectrum, where enumerating closed polygons scales as $\mathcal{O}(N^{p-1})$,
but the FFT remains $\mathcal{O}(pN\log N)$.

\begin{figure}
  \centering
  \includegraphics[width=0.45\textwidth]{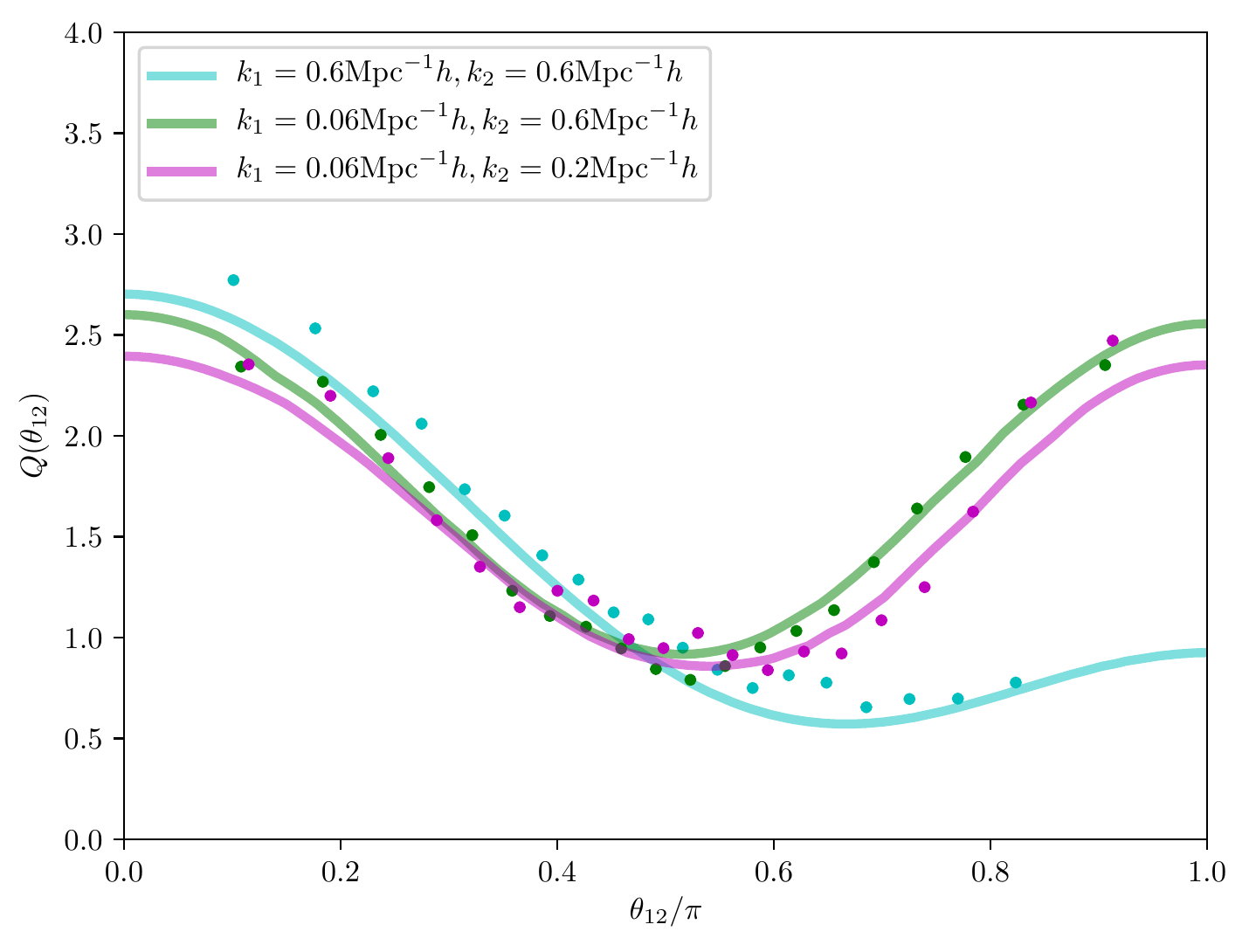}
  \caption{The reduced bispectrum $Q(k_1,k_2,\theta)$ for a matter overdensity
    field generated from an $N$-body simulation at $z = 7$. The points
    correspond to the points from our estimator, and the solid lines correspond
    to the reduced analytic expression in Equation~(\ref{eqn:matter_bspec}). The
    points are slightly offset from each other in the $x$-direction for visual
    clarity.}
  \label{fig:qk}
\end{figure}

In order to validate the result of the estimator, we compute the bispectrum of
the matter density field from an $N$-body simulation. \citet{fry1984} gives an
expression for the bispectrum using second-order perturbation theory:
\begin{equation}
B(\vb{k}_1,\vb{k}_2,\vb{k}_3) = 2F(\vb{k}_1,\vb{k}_2)P(k_1)P(k_2) + (\mathrm{cyc.}),
\label{eqn:matter_bspec}
\end{equation}
where $F(\vb{k}_1,\vb{k}_2)$ is defined as \citep{scoccimarro2000}:
\begin{equation}
F(\vb{k}_1,\vb{k}_2) = \frac{5}{7} + \qty(\frac{\vb{k}_1 \cdot \vb{k}_2}{2k_1 k_2})\qty(\frac{k_1}{k_2} + \frac{k_2}{k_1})
  + \frac{2}{7} \qty(\frac{\vb{k}_1 \cdot \vb{k}_2}{k_1 k_2})^2.
\label{eqn:ffunc}
\end{equation}
A related quantity that allows for readily comparing the results of the
bispectrum from different spatial scales is the reduced bispectrum
$Q(k_1,k_2,\theta_{12})$, defined in Equation~(\ref{eqn:reduced_bspec}).

Figure~\ref{fig:qk} shows the reduced bispectrum $Q(k_1,k_2,\theta_{12})$ as
computed using our direct estimator compared with an analytic expression based
on Equation~(\ref{eqn:matter_bspec}) for several different combinations of
$k_1$ and $k_2$. The matter density field is a snapshot from an $N$-body
simulation at $z = 7$. As can be seen, the agreement between the estimator and
the analytic result is broadly consistent. These results are qualitatively
similar to those from \citet{watkinson_etal2017}.

\bibliography{/Users/plaplant/Dropbox/school/Penn/research/bibliography/mybib}
\bibliographystyle{aasjournal}

\end{document}